\documentclass{emulateapj}
\usepackage{wasysym}
\usepackage{graphicx}
\usepackage{amssymb}
\usepackage{textcomp}

\def\ltsima{$\; \buildrel < \over \sim \;$}
\def\lsim{\lower.5ex\hbox{\ltsima}}
\def\gtsima{$\; \buildrel > \over \sim \;$}
\def\gsim{\lower.5ex\hbox{\gtsima}}

\shorttitle{RM effect for WASP-1 and WASP-2}
\shortauthors{Albrecht et al.}
\begin{document}

\title{
  Two Upper Limits on the Rossiter-McLaughlin Effect, with Differing
  Implications: \\
  WASP-1 has a High Obliquity and WASP-2 is Indeterminate
  \altaffilmark{$\star$}
}

\author{
  Simon Albrecht\altaffilmark{1}, 
  Joshua N.\ Winn\altaffilmark{1},
  John Asher Johnson\altaffilmark{2},
  R.\ Paul Butler\altaffilmark{3},
  Jeffrey D.\ Crane\altaffilmark{4},
  Stephen A.\ Shectman\altaffilmark{4},
  Ian B.\ Thompson\altaffilmark{4},
  Norio Narita\altaffilmark{5},
  Bun'ei Sato\altaffilmark{6},
  Teruyuki Hirano \altaffilmark{7,1},
  Keigo Enya\altaffilmark{8},
  Debra Fischer\altaffilmark{9} 
}

\altaffiltext{1}{Department of Physics, and Kavli Institute for
  Astrophysics and Space Research, Massachusetts Institute of
  Technology, Cambridge, MA 02139, USA}

\altaffiltext{2}{California Institute of Technology, Department of
  Astrophysics, MC249-17, Pasadena, CA 91125; NASA Exoplanet Science
  Institute (NExScI), USA}

\altaffiltext{3}{Department of Terrestrial Magnetism, Carnegie
  Institution of Washington, 5241 Broad Branch Road NW, Washington, DC
  20015, USA}
 
\altaffiltext{4}{The Observatories of the Carnegie Institution of
  Washington, 813 Santa Barbara Street, Pasadena, CA 91101, USA}

\altaffiltext{5}{National Astronomical Observatory of Japan, 2-21-1
  Osawa, Mitaka, Tokyo, 181-8588, Japan}
 
\altaffiltext{6}{Department of Earth and Planetary Sciences, Graduate
  School of Science and Engineering, Tokyo Institute of Technology,
  2-12-1 Ookayama, Meguro-ku, Tokyo 152-8551, Japan}

\altaffiltext{7}{Department of Physics, The University of Tokyo,
 Tokyo 113-0033, Japan}

\altaffiltext{8}{Department of Infrared Astrophysics, Institute of
  Space and Astronautical Science, Japan Aerospace Exploration Agency,
  3-1-1, Yoshinodai, Chuo-ku, Sagamihara, Kanagawa 252-5210, Japan}

\altaffiltext{9}{Department of Astronomy, Yale University, New Haven,
  CT 06511, USA}

\altaffiltext{$\star$}{
  The data presented herein were collected with the the Magellan
  (Clay) Telescope located at Las Campanas Observatory, Chile; the
  Subaru telescope, which is operated by the National Astronomical
  Observatory of Japan; and the Keck~I telescope at the W.M.\ Keck
  Observatory, which is operated as a scientific partnership among the
  California Institute of Technology, the University of California and
  the National Aeronautics and Space Administration.
}

\begin{abstract}

  We present precise radial-velocity measurements of WASP-1 and WASP-2
  throughout transits of their giant planets. Our goal was to detect
  the Rossiter-McLaughlin (RM) effect, the anomalous radial velocity
  observed during eclipses of rotating stars, which can be used to
  study the obliquities of planet-hosting stars. For WASP-1 a weak
  signal of a prograde orbit was detected with $\approx$2$\sigma$
  confidence, and for WASP-2 no signal was detected. The resulting
  upper bounds on the RM amplitude have different implications for
  these two systems, because of the contrasting transit geometries and
  the stellar types. Because WASP-1 is an F7V star, and such stars are
  typically rapid rotators, the most probable reason for the
  suppression of the RM effect is that the star is viewed nearly
  pole-on. This implies the WASP-1 star has a high obliquity with
  respect to the edge-on planetary orbit. Because WASP-2 is a K1V
  star, and is expected to be a slow rotator, no firm conclusion can
  be drawn about the stellar obliquity. Our data and our analysis
  contradict an earlier claim that WASP-2b has a retrograde orbit,
  thereby revoking this system's status as an exception to the pattern
  that cool stars have low obliquities.

 \end{abstract}

\keywords{techniques: spectroscopic  -- stars: rotation -- planetary systems  -- planets and satellites:
  formation -- planet-star interactions}

\section{Introduction}
\label{sect:introduction}

The existence of Jupiter-sized planets on very close-in orbits
presents a challenge to any model which aims to explain the formation
of planets. In the current picture, these planets form further away
from their host star and migrate inward. How and why this migration
occurs is subject to debate \cite[e.g.][]{lin1996,nagasawa2008}.
Recently an important clue to this riddle was revealed: a subset of
the close-in planets have orbits that are seeming randomly-oriented
with respect to the equatorial plane of the host star (see, e.g.,
\citealt{hebrard2008,winn2009,narita2009,johnson2009,triaud2010}).

\cite{winn2010} and \cite{schlaufman2010} found that planets orbiting
stars with effective temperatures $\gsim$6250~K (i.e., mass
  $\gsim$1.2$M_\odot$) tend to have an orbital axis misaligned with
respect to the stellar spin axis, i.e., a high stellar obliquity. In
contrast, the two axes are generally well-aligned for systems in which
the host star is cooler (i.e., less massive). These authors
noted that this could reflect a difference in the dominant planet
migration mechanism between low-mass stars and high-mass stars.
\cite{winn2010} further speculated that {\it all} close-in giant
planets are transported inwards by processes that disrupt spin-orbit
alignment. Subsequently, the angular momenta are realigned via tidal
interaction, and this process is more rapid in cooler stars perhaps
due to their thicker convective envelopes. In this picture any viable
migration process would have to introduce misalignment between orbital
and stellar spin.

However, the small sample of accurate and precise measurements of
stellar obliquities ($\approx$$25$ systems) and the possibility of
selection effects present us with many pitfalls if we want to validate
or reject theories of giant planet migration. Here we report on our
attempts to measure the spin-orbit angles in the WASP-1 and WASP-2
systems, taking advantage of the Rossiter-McLaughlin (RM) effect. \\

\begin{figure*}
  \begin{center}
   \includegraphics[width=17.cm]{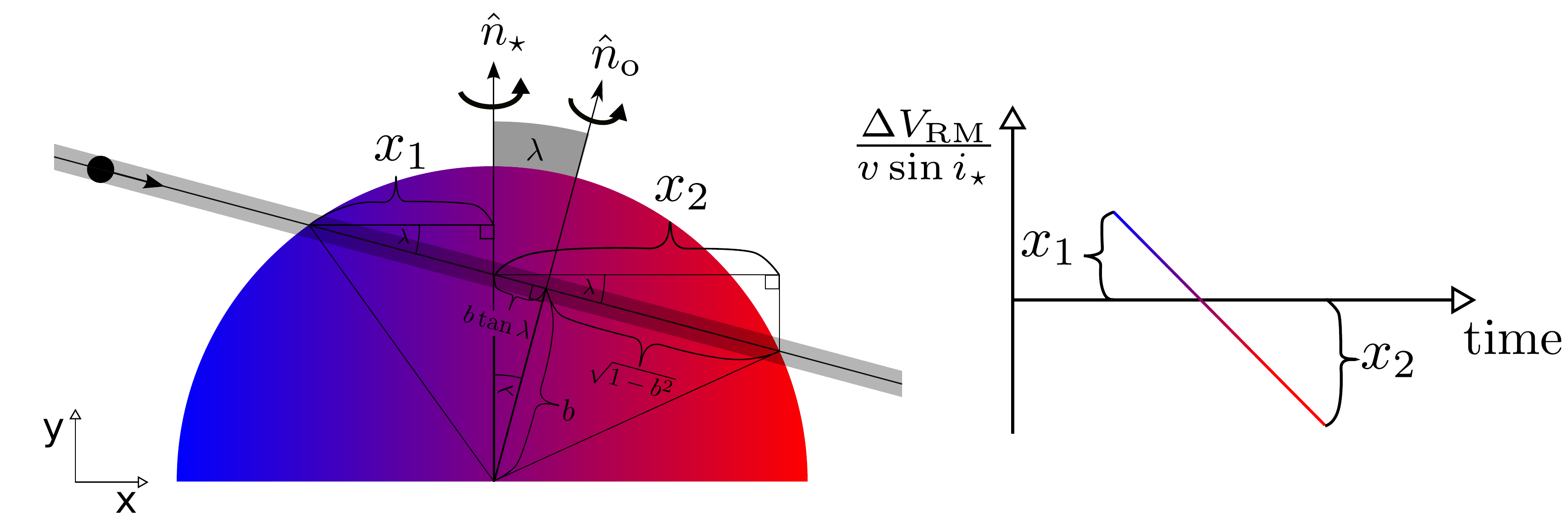}
   \caption {\label{fig:rm} {\bf Geometry of the Rossiter-McLaughlin
       effect.} The left panel illustrates a transit, with the planet
     crossing from left to right. Due to stellar rotation the left
     side of the star is moving towards the observer and the right
     side is receding. The unit vectors $\hat{n}_\star$ and
     $\hat{n}_{\rm o}$ point along the sky-projected stellar rotation
     axis and planetary orbital axis. They are separated by an angle
     $\lambda$. In this diagram, $\hat{n}_\star$ points in the
     $y$-direction, and the anomalous radial-velocity caused by the
     planet is proportional to $x$ (see Section 2). The extrema in the
     RM signal occur at ingress ($x=x_1$) and egress ($x=x_2$). The
     relations between $x_1$, $x_2$, $\lambda$ and the impact
     parameter $b$ are indicated on the diagram. The right panel shows
     the corresponding RM signal as a function of time, for an
       idealized case with no stellar limb darkening.}
  \end{center}
\end{figure*}

\noindent {\it --- WASP-1b} was discovered by \cite{cameron2007}.  It
orbits on a $2\fd52$ circular orbit around a F7V star and has a mass
of 0.92\,M$_{\rm Jup}$. One reason why this system is interesting is
that \cite{stempels2007} reported a projected stellar rotation speed
of $v\sin i_\star < 5.79\pm0.35$ km\,s$^{-1}$, which is relatively
slow for a star of this spectral type. For this reason,
\citet{schlaufman2010} identified WASP-1 as a likely case of
spin-orbit misalignment along the line of sight, i.e., $\sin i_\star <
1$ even though $\sin i_{\rm o}\approx 1$ for the planetary orbit. The
star's effective temperature places it right in the range where the
transition from well-aligned to misaligned orbits was observed by
\citet{winn2010} and \citet{schlaufman2010}. Recently
\cite{simpson2011} reported a detection of the RM effect for this
system and concluded the orbital and stellar spins were misaligned in
the plane of the sky. As we will discuss in Section \ref{sect:wasp1},
our analysis leads to a more complex conclusion: while we agree that
the spin and orbital vectors are misaligned, the evidence for a
sky-plane misalignment is much weaker than the evidence
for a line-of-sight misalignment.\\

\noindent {\it --- WASP-2b} was also discovered by \cite{cameron2007}.
This 0.87\,M$_{\rm Jup}$ planet has a host star of later spectral
type (K1V) and orbits on a circular $2\fd15$ orbit. Recently
\cite{triaud2010} reported an angle of $153^{+11}_{-15}$ degrees
between the projected orbital and stellar spins, i.e., a retrograde
orbit. This is interesting as the host star is firmly on the ``cool''
side of the proposed divide between cool well-aligned stars and hot
misaligned stars. WASP-2 would therefore constitute an important
exception to the trend. However, as we will discuss in
Section~\ref{sect:wasp2}, we find no evidence for a retrograde orbit
and argue that the obliquity of the host star cannot be determined
from either the new data or the previously published data.

\section{Rossiter-McLaughlin effect}
\label{sect:rm_effect}

From the perspective of this study there are two main differences
between the WASP-1 and WASP-2 systems. First, the stars are of
differing spectral type, leading to different {\it a priori}
expectations for the stellar rotation speed. The implications of this
difference are discussed in Sections~\ref{sect:wasp1}
and~\ref{sect:wasp2}. Second, the planets' trajectories across the
stellar disk have different impact parameters: WASP-1b nearly crosses
the center of the disk, while the transit of WASP-2b is
off-center. This section is concerned with the implications of this
geometrical difference, as well as the more general relation between
the characteristics of the RM signal and the parameters that are often
used to model the signal. Some of these aspects of RM modeling were
described by \cite{gaudi2007}, to which we refer the reader for a more
comprehensive account.

Models of the RM effect with varying degrees of accuracy have been
worked out by \citet{hosokawa1953, queloz2000, otha2005, winn2005,
  gimenez2006, albrecht2007, gaudi2007, cameron2010,hirano2010} and
\citet{shporer2011}. Because our aim in this section is pedagogical,
we ignore the influence of stellar limb-darkening,
differential rotation, gravity darkening, surface velocity fields and
any departures from sphericity of the planet or star. We also
assume that the planet-to-star radius ratio $R_p/R_\star$ is small,
and that this parameter is known precisely along with all the other
parameters that are derived from photometric observations of
transits. In particular we assume precise knowledge of the impact
parameter $b \equiv r_t\cos i_o/R_\star$, where $r_t$ is the orbital
distance at the time of transit, $R_{\star}$ is the stellar radius,
and $i_{\rm o}$ is the orbital inclination.

With these approximations, the anomalous radial velocity due to the RM
effect is
\begin{equation}
\Delta V_{\rm RM}(t) \approx -\left( \frac{R_p}{R_\star} \right)^2 v_{\rm p}(t),
\end{equation}
where $v_p(t)$ is the ``subplanet'' radial velocity, i.e., the radial
component of the rotational velocity of the portion of the photosphere
hidden by the planet. Neglecting differential rotation, we may
write
\begin{equation}
v_{\rm p}(t) = (v\sin i_\star)~x/R_\star,
\end{equation}
where $x$ is the distance on the sky plane from the center of the
planet to the stellar rotation axis [see, e.g., pages 461-462 of \cite{gray2005}].

\begin{figure}
  \begin{center}
   \includegraphics[width=8.cm]{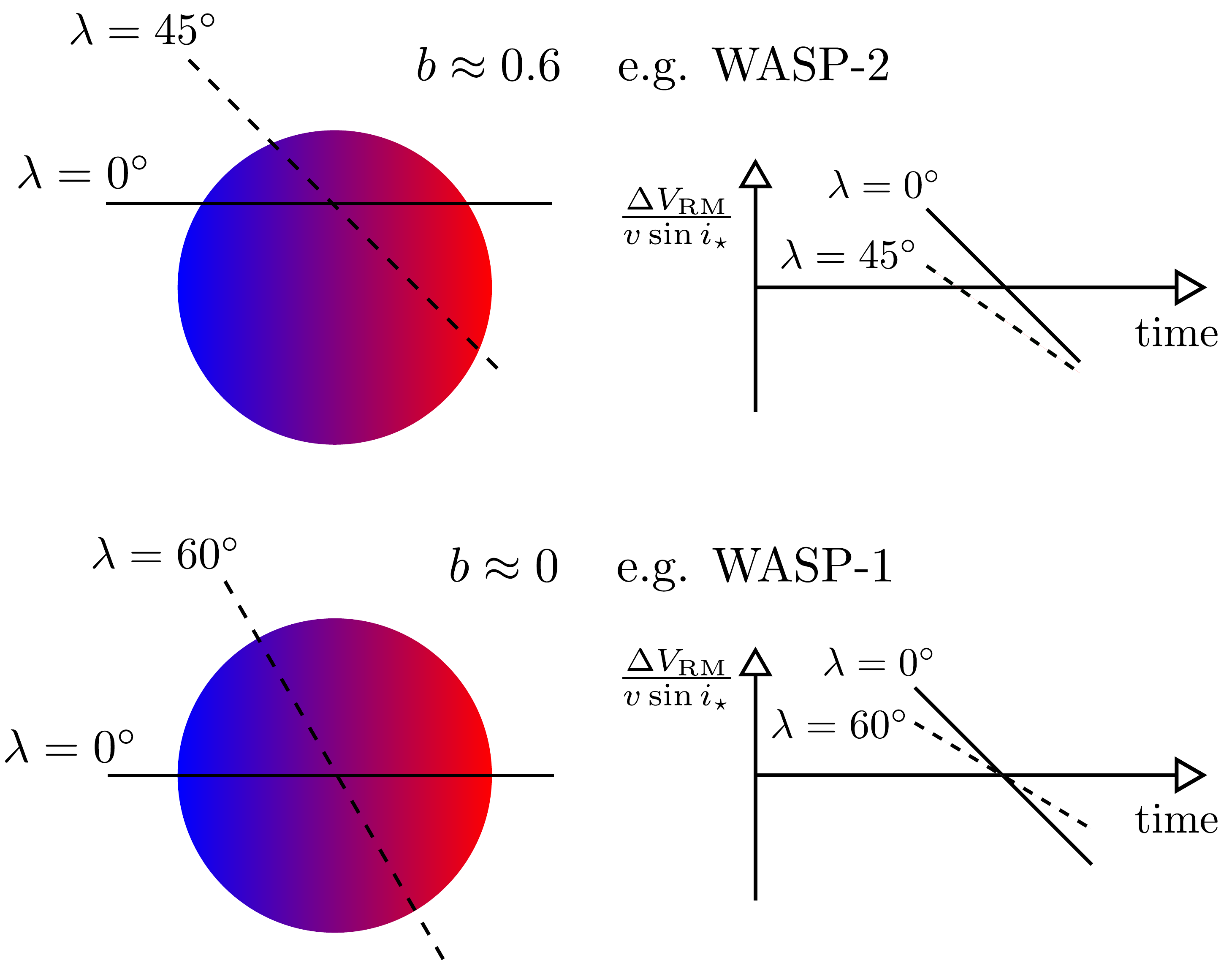}
   \caption {\label{fig:rm_ex} {\bf The dependence of the RM signal on
       $\lambda$, for high and low impact parameters.} The upper left
     panel shows the geometry for a system with $b\approx 0.6$, for
     two different cases of $\lambda$. In the first case (solid line)
     the orbital and stellar spins are aligned, and in the second case
     (dashed line) they are misaligned. The upper right panel shows
     the corresponding RM signals; both the mean amplitude and the
     asymmetry of the RM signal are different. The two lower panels
     show a similar orbital configurations but for $b\approx 0$.
     Here, the mean amplitude changes with $\lambda$ but the asymmetry
     is always zero. }
  \end{center}
\end{figure}

The situation is illustrated in Figure~\ref{fig:rm}. In this diagram,
$\hat{n}_\star$ and $\hat{n}_{\rm o}$ are unit vectors parallel to the
sky projections of the stellar and orbital angular momenta,
respectively. The angle $\lambda$ is measured from $\hat{n}_\star$ to
$\hat{n}_{\rm o}$.\footnote{This definition of $\lambda$ is taken from
  \cite{otha2005}. Some other investigators measure the angle from
  $\hat{n}_{\rm o}$ to $\hat{n}_\star$ and denote the angle $\beta$. Clearly
  $\beta=-\lambda$.} The maximum redshift and blueshift occur at
ingress and egress, which we take to have $x$-coordinates of $x_{1}$
and $x_{2}$ respectively. Using the geometrical relations shown in the
diagram, we may write $x_1$ and $x_2$ in terms of $b$ and $\lambda$,
{\small
\begin{eqnarray}
\label{eqn:x}
x_{1} & = & \left( \sqrt{1-b^{2}} - b \tan \lambda \right) \cos\lambda
= \sqrt{1-b^{2}}\cos\lambda - b \sin \lambda,\,\,\,\,\,\,\nonumber \\
x_{2} & = & \left( \sqrt{1-b^{2}} + b \tan  \lambda \right) \cos\lambda 
= \sqrt{1-b^{2}}\cos\lambda + b \sin \lambda.\,\,\,\,\,\,
\end{eqnarray}
}
It is instructive to examine the (scaled) sum and difference of $x_1$
and $x_2$,
\begin{eqnarray}
\frac{1}{2}~v\sin i_\star~(x_2 + x_1) & = & \sqrt{1-b^2}~v\sin i_\star\cos\lambda, \nonumber \\
\frac{1}{2}~v\sin i_\star~(x_2 - x_1) & = & b~v\sin i_\star\sin\lambda.
\end{eqnarray}
The sum is the {\it mean amplitude} of the red and blue peaks of the
RM effect, while the difference is a measure of {\it asymmetry}
between the peaks. For a fixed $b$, the mean amplitude depends on
$v\sin i_\star \cos\lambda$ while the asymmetry depends on $v\sin
i_\star \sin\lambda$.

Figure~\ref{fig:rm_ex} shows the RM signal in 4 different situations:
two different values of $\lambda$ for each of two different impact
parameters. The upper panels show the case $b\approx 0.6$, as is the
case for WASP-2. Here, as $\lambda$ is varied, both the mean amplitude
and asymmetry of the RM signal are observed to change. By measuring
the mean amplitude and asymmetry, one may determine both $v\sin
i_\star$ and $\lambda$. The lower panels show the case $b \approx 0$,
as is the case for WASP-1. Here, the asymmetry vanishes regardless of
$\lambda$. The only observable quantity is the mean amplitude, and
therefore the only parameter combination that can be determined is
$v\sin i_\star \cos\lambda$.

Consequently, for transits with low impact parameters, $\lambda$ and
$v\sin i_\star$ have strongly correlated uncertainties and it is not
possible to measure $\lambda$ without some prior information about
$v\sin i_\star$. However, in such cases it is still possible to tell
whether $\cos\lambda$ is positive or negative, and therefore whether
the orbit is prograde ($|\lambda| < 90^\circ$) or retrograde
($|\lambda| > 90^\circ$). We also note that the degeneracy between
$v\sin i_\star$ and $\lambda$ can be broken in principle when the RM
effect is modeled at the level of spectral-line distortion, rather
than modeling only the anomalous radial velocity \citep{albrecht2007,
  cameron2010}. In this paper, though, we work with the anomalous
radial velocity.

\section{ WASP-1}
\label{sect:wasp1}

\subsection{Observations and basic stellar parameters}
\label{sect:wasp1_obs}

\begin{figure}
  \begin{center} 
 \includegraphics[width=8cm]{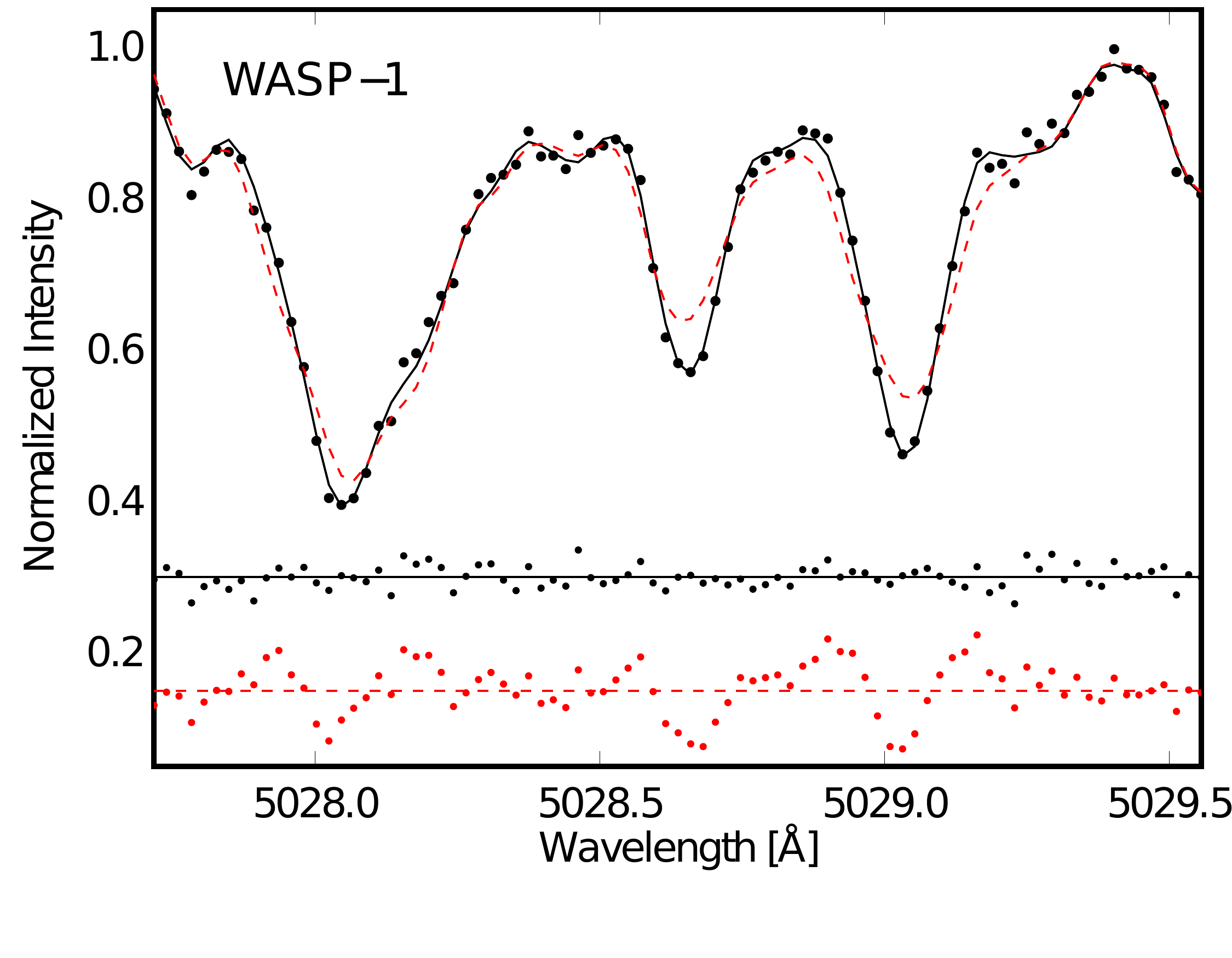}
 \caption {\label{fig:wasp1_spec} {\bf Spectrum of the F7V star
     WASP-1.} A small portion of the the spectrum of WASP-1, as
   obtained with HIRES, is shown. The dots represent the the observed
   spectrum, the solid line represents our best fit with a
   macro-turbulence parameter of $3.98$~km\,s$^{-1}$ and a $v\sin
   i_\star$ of $2.9$~km~s$^{-1}$. The (red) dashed line shows the
   spectrum broadened to the values given by \cite{stempels2007}, who
   obtained a $v\sin i_\star$ of $5.79$~km~s$^{-1}$ with a
   macro-turbulence parameter of 4.5~km\,s$^{-1}$.  The two lower rows
   of points show the differences between model and data for our best
   fit (black dots) and the values given by \cite{stempels2007} (red dots).}
  \end{center}
\end{figure}

We conducted spectroscopic observations of WASP-1 transits with the
Keck\,I 10\,m telescope and the Subaru 8.2\,m telescope.  With Keck,
we used the High Resolution Spectrograph (HIRES; \citealt{vogt1994})
to gather 34 spectra spanning the transit of 2007~September~1/2. With
Subaru, we used the High Dispersion Spectrograph (HDS;
\citealt{noguchi2002}) to observe two different transits, on the
nights of 2007 August 4/5 and 2007 September 6/7. A total of 23
spectra were obtained with HDS, most of which (20) were obtained on
the latter night. At both observatories an iodine gas absorption cell
was used to correct for changes in the point spread function and
wavelength scale. Radial velocities (RV) were derived from the spectra
using procedures similar to those described by \cite{butler1996}. See
\cite{sato2002} and \cite{narita2007} for details on the Subaru data
reduction. The RVs are shown in Figure~\ref{fig:wasp1_rv} and given in
Table~\ref{tab:wasp1_rv}.

To check on the basic stellar parameters, we used the Spectroscopy
Made Easy (SME) software package \citep{valenti1996} to model the
high-resolution, high--signal-to-noise ratio template spectrum. We
obtained $T_{\rm eff} = 6213\pm 51$~K, $\log g=4.19\pm 0.07$, $[M/H] =
0.17\pm 0.05$, and $v\sin i_\star = 1.60\pm 0.50$~km~s$^{-1}$. These
can be compared to the previous spectroscopic results of WASP-1 by
\citet{stempels2007}, which gave $T_{\rm eff} = 6110\pm 45$~K, $\log g
= 4.28\pm 0.15$, $[M/H] = 0.23\pm 0.08$, and $v\sin i_\star = 5.79\pm
0.35$~km~s$^{-1}$. Our analysis gave a higher value of
$T_{\rm eff}$ and a lower value of $v\sin i_\star$.

\begin{figure}
  \begin{center}
   \includegraphics[width=8cm]{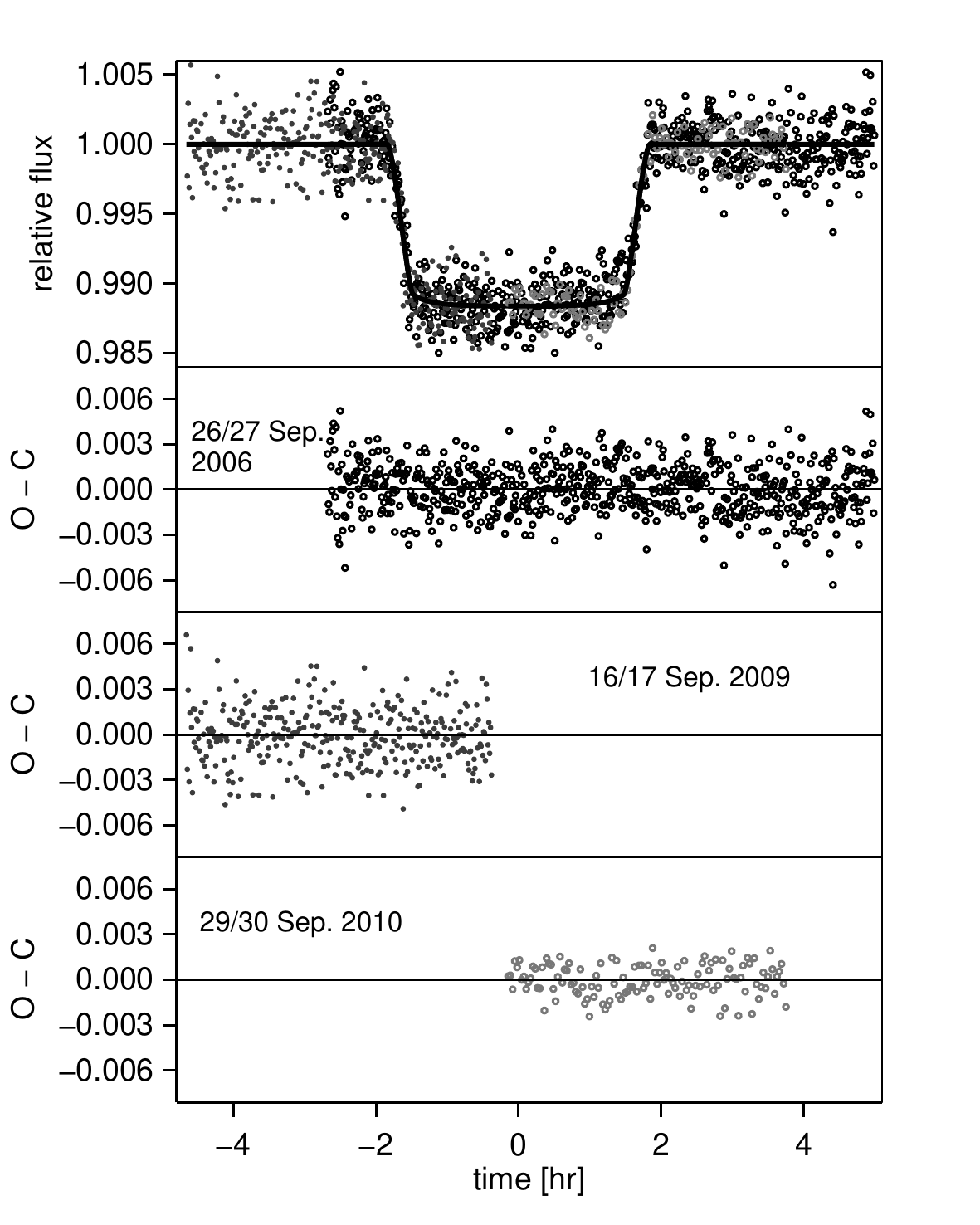}
   \caption {\label{fig:wasp1_phot} {\bf Photometry of WASP-1
       transits.} The upper panel is a composite $z'$-band light curve
     based on our data and that of \citet{charbonneau2007}.  The lower
     three panels show the residuals between each of the 3 datasets
     and the best-fitting model.}
  \end{center}
\end{figure}

The discrepancy in $T_{\rm eff}$ is discussed in
Section~\ref{sect:discussion}. The discrepancy in $v \sin i_\star$ is
of immediate importance because stellar rotation is a key parameter in
the interpretation of the RM effect. Frequently, such discrepancies
arise because of differing assumptions regarding turbulent
broadening. SME determines $v\sin i_\star$ based on the observed
widths of numerous weak lines in the spectrum. The widths
are influenced not only by rotation, but also by random motions of the
stellar photosphere (microturbulence and macroturbulence), and these
effects cannot generally be disentangled. Hence it is necessary to
assume ``typical'' values of the turbulence parameters and attribute
the excess broadening of the observed lines to rotation. When using
SME, it is assumed $v_{\rm mic} = 0.85$~km~s$^{-1}$ and
\begin{equation}
\label{eq:macro}
v_{\rm mac} = \left( 3.98 + \frac{T_{\rm eff} - 5770~{\rm K}}{650~{\rm
    K}} \right)~{\rm km~s}^{-1},
\end{equation}
an empirical relation determined by \citet{valenti2005}.\footnote{The
  equation given here corrects a sign error in Equation (1) of
  \citet{valenti2005}.} For WASP-1, this formula gives $v_{\rm mac} =
4.66$~km~s$^{-1}$. This is not too different from the value $v_{\rm
  mac} = 4.5$~km~s$^{-1}$ that was assumed by \cite{stempels2007} and
hence the discrepancy in $v\sin i_\star$ cannot be attributed to
different assumptions regarding macroturbulence.

\begin{figure} 
  \begin{center} 
 \includegraphics[width=8.cm]{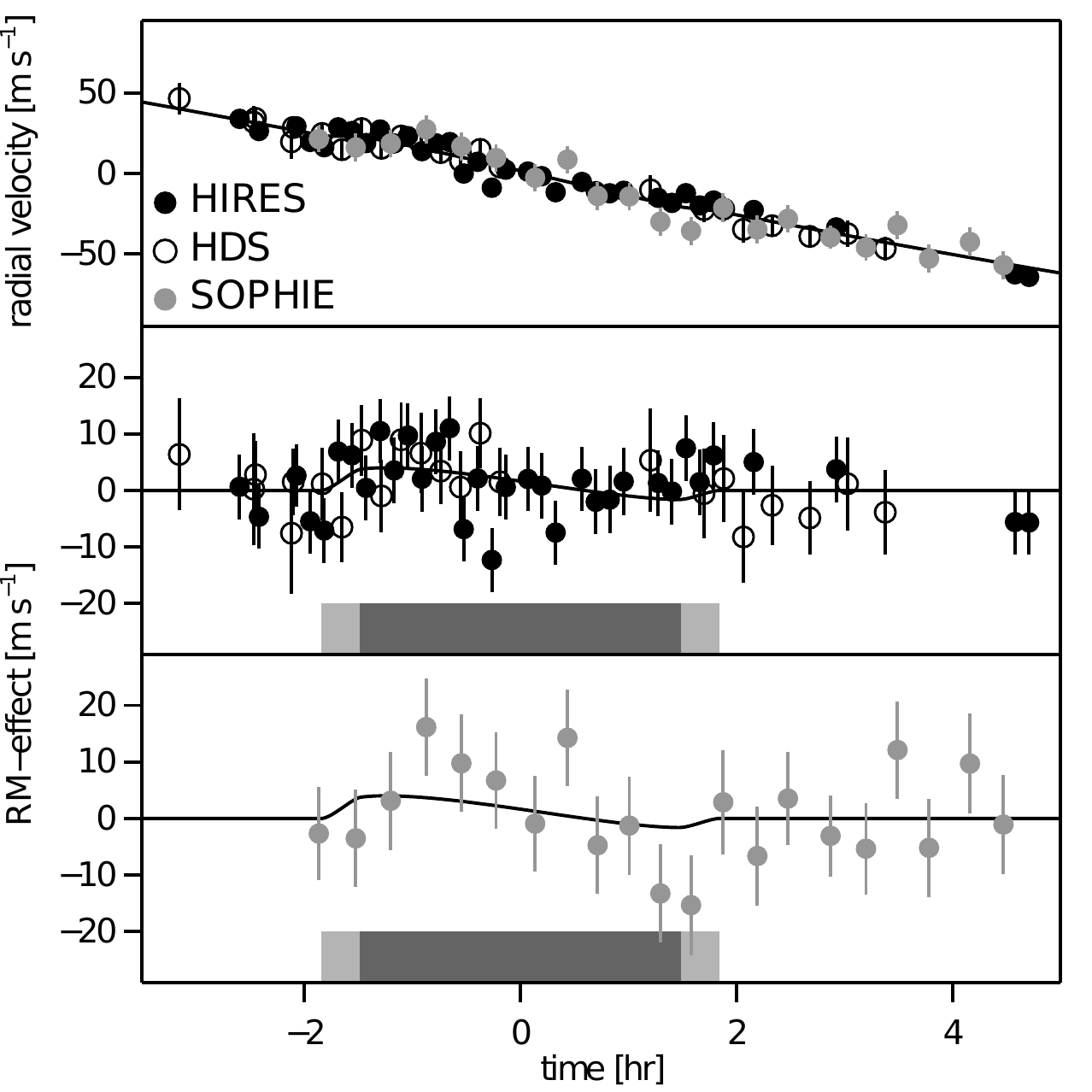}
 \caption {\label{fig:wasp1_rv} {\bf Spectroscopy of WASP-1
     transits. }  The radial velocities measured before, during, and
   after transit are plotted as a function of time from inferior
   conjunction. Solid symbols are data from HIRES and open symbols are
   data from HDS. Gray symbols are the SOPHIE data from
   \cite{simpson2011}, which are shown for comparison only (they were
   not used in our fitting process). The upper panel shows the
   measured RVs and the best-fitting model. In the middle panel, the
   orbital contribution to the observed RVs has been subtracted,
   isolating the RM effect. The lower panel shows the SOPHIE RVs after
   subtracting our best-fitting orbital model. The light and dark gray
   bars in the two lower panels indicate times of first, second, third,
   and fourth contact.}
  \end{center}
\end{figure}

To investigate further, we performed a differential assay for
rotation, based on a comparison between the Solar spectrum and a
Keck/HIRES spectrum of WASP-1. First, we deconvolved the WASP-1
spectrum to remove the instrumental broadening of width
2.2~km~s$^{-1}$. Then, using the MORPH code of \cite{johnson2006}, we
applied a rotational broadening kernel to the NSO Solar spectrum of
\cite{kurucz1984} to achieve the best fit to the deconvolved WASP-1
spectrum. We found that the best-fitting broadening kernel was
2.36~km~s$^{-1}$, indicating the WASP-1 lines are slightly broader
than the Solar lines. Figure~\ref{fig:wasp1_spec} shows a small
portion of the WASP-1 spectrum and our best-fitting model based on the
broadened Solar spectrum.

The larger breadth of the WASP-1 lines could be interpreted as more
rapid rotation than the Sun, but in fact part of the increased breath
is expected to be due to the higher macroturbulence of WASP-1. However
since the accuracy of Eqn.~\ref{eq:macro} is not known, we may here
simply assume that the macroturbulence of WASP-1 is greater than or
equal to the macroturbulence of the Sun. The MORPH finding implies
\begin{eqnarray}
\label{eq:morph}
& & [v\sin i_\star~({\rm W1})]^2 + [v_{\rm mac}~(\odot)]^2
\approx \nonumber \\
& & [v\sin i_\star~(\odot)]^2 + [v_{\rm mac}~(\odot)]^2 +
(2.36~{\rm km~s}^{-1})^2,
\end{eqnarray}
where the ``W1'' quantity is for WASP-1 and the ``$\odot$''
quantities are for the Sun.\footnote{We verified with numerical
  experiments that in this regime of velocity widths and for the SNR
  and resolution of our spectrum, the widths of the various
  convolution kernels can be approximately added in quadrature as
  implied here.} Taking the disk-integrated rotation and
macroturbulence of the Sun to be 1.63~km~s$^{-1}$ and
3.98~km~s$^{-1}$, and the macroturbulence for WASP-1 the same as the
sun, Equation~\ref{eq:morph} gives $v\sin i_\star < 2.9$~km~s$^{-1}$
for WASP-1. 

\begin{figure} 
  \begin{center} 
 \includegraphics[width=8.cm]{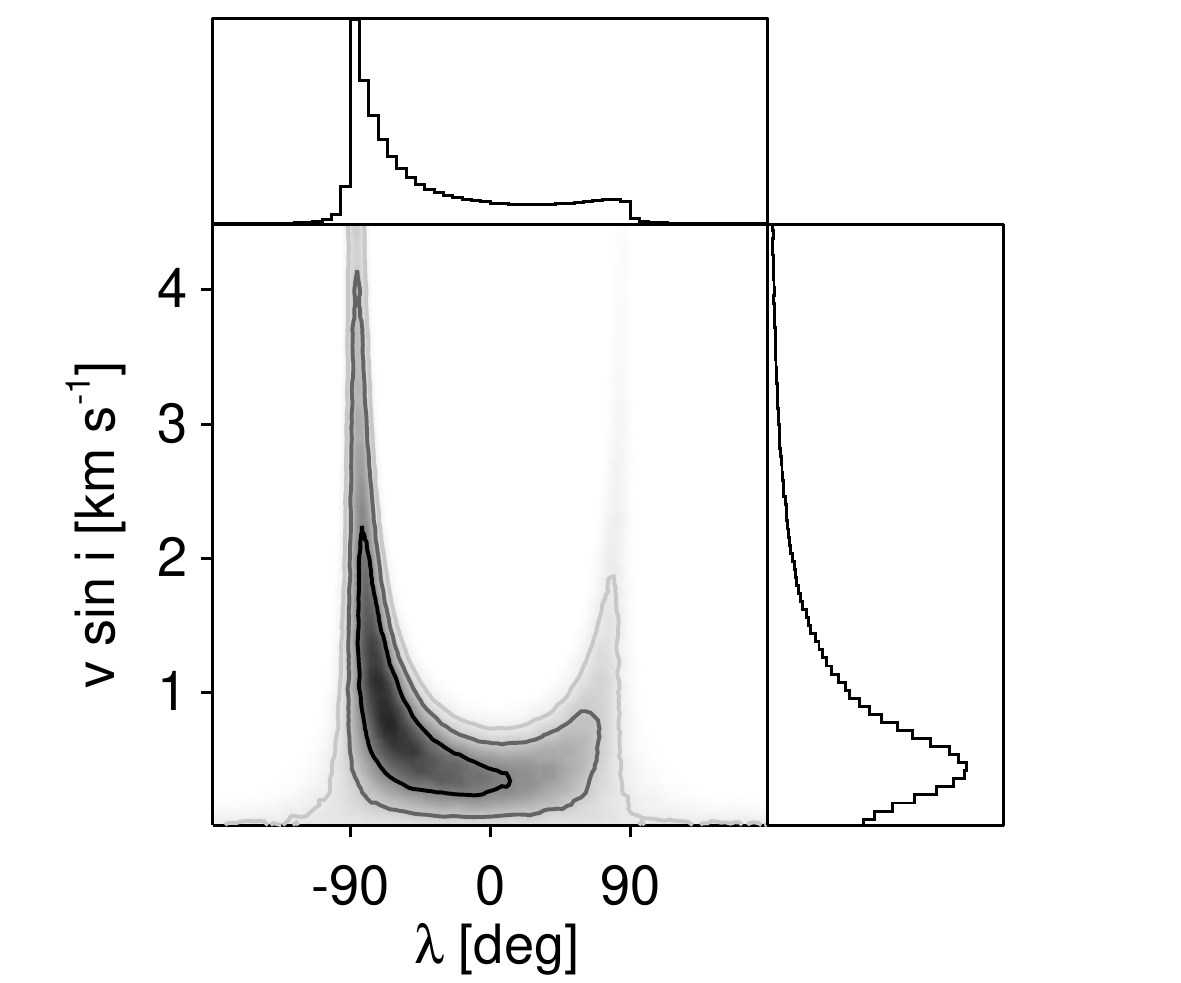}
 \caption {\label{fig:wasp1_mcmc} {\bf Results for $v \sin i_{\star}$
     and $\lambda$, based on our MCMC analysis in the WASP-1 system. }
   The gray scale indicates the posterior probability density,
   marginalized over all other parameters. The contours represent the
   2-D 68.3\%, 95\%, and 99.73\% confidence limits. The
   one-dimensional marginalized distributions are shown on the sides
   of the contour plot. A strong correlation between the projected
   rotation speed and the projected angle between the stellar and
   orbital spins exits. Either the two axes are nearly perpendicular
   on the sky plane, or else $v\sin i_\star$ is small and $\lambda$
   can have any value.}
  \end{center}
\end{figure}

These results show that the projected rotation speed of WASP-1 is
quite slow ($<$2.9~km~s$^{-1}$) and is in fact nearly undetectable
against the dominant line-broadening effect of
macroturbulence. Figure~\ref{fig:wasp1_spec} also shows that our
spectrum is incompatible with the more rapid rotation of $5.79\pm
0.35$~km~s$^{-1}$ found by \cite{stempels2007}. We do not know why
\cite{stempels2007} found a higher $v\sin i_\star$ even when making
equivalent assumptions regarding macroturbulence. Genuine changes in
$v\sin i_\star$ could be produced by spin precession, but are not
expected to be appreciable on such short timescales, and hence we
proceed under the assumption that the \cite{stempels2007}
determination was in error.

To reduce the uncertainties in the photometric parameters we gathered
new photometric data with Keplercam, a CCD camera on the 1.2\,m
telescope of the Fred L. Whipple Observatory on Mount Hopkins, Arizona
\citep{szentgyorgyi2005}. Observations were conducted in the SDSS
$z'$-band on 2009~September~16/17 and 2010~September~29/30, although
bad weather interrupted the transit in both cases. The new photometric
data were combined with the previous data of \citet{charbonneau2007},
which were gathered with the same instrument and reduced with similar
procedures.  All of the Keplercam data are shown in
Figure~\ref{fig:wasp1_phot}.

\begin{figure}
  \begin{center} 
    \includegraphics[width=8.cm]{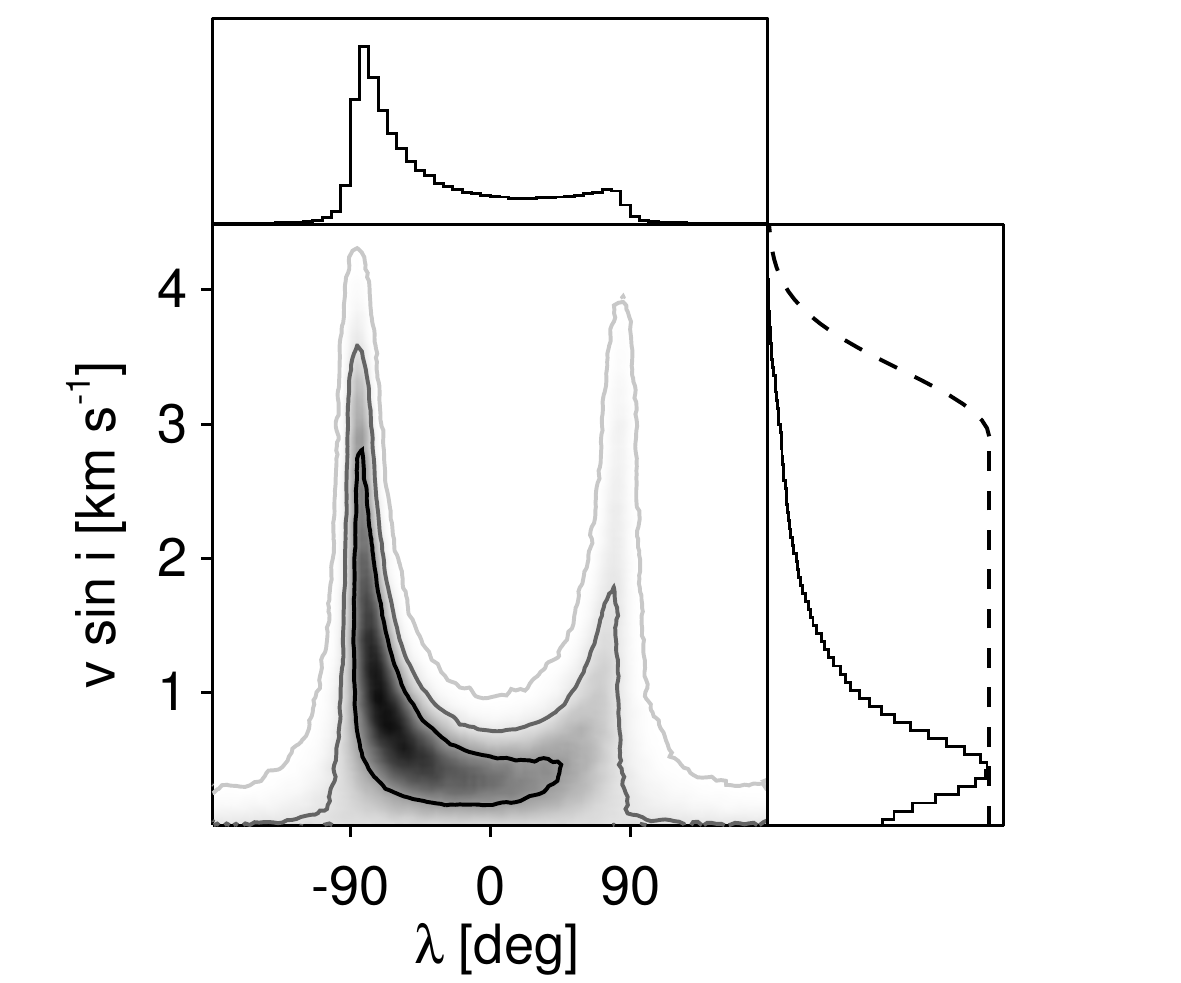}
    \caption {\label{fig:wasp1_mcmc_vsini} {\bf Results for $v\sin
        i_\star$ and $\lambda$, this time including a prior constraint
        on $v\sin i_\star$. }The prior constraint was based on the
      spectroscopic result $v\sin i_\star < 2.9$~km~s$^{-1}$ (see
      Section 3.1) and is illustrated by the dashed line in the
      right-hand side panel. For $v\sin i_\star < 2.9$~km~s$^{-1}$,
      the prior was set equal to unity; and for greater values the
      prior was a Gaussian function with mean 2.9~km~s$^{-1}$ and
      standard deviation 0.5~km~s$^{-1}$.  Compared to Figure~
      \ref{fig:wasp1_mcmc}, the solutions are similar but are
      constrained to have somewhat lower $v\sin i_\star$.}
  \end{center}
\end{figure}

\subsection{Analysis}

To derive constraints on $\lambda$, we fitted a model simultaneously
to the RV data and the photometric data. The photometric transit was
modeled with the code of \cite{mandel2002}, and the RM effect was
modeled with a simplified version of the code of \cite{albrecht2007}.
This model for the RM effect is similar to that given in
Eqn.~\ref{eqn:stat_wasp1} but takes limb darkening into account. It
does not take into account the nonlinear relation between $\Delta
V_{\rm RM}$ and $v_p(t)$ because those nonlinearities are important
only for stars with larger $v\sin i_\star$ \cite[see,
e.g.][]{winn2005,hirano2010}.

The transit impact parameter for WASP-1b is small, with
\cite{torres2008} having reported $b=0.00^{+0.27}_{-0.00}$.
Therefore, based on the reasoning of Section~\ref{sect:rm_effect}, we
expect the data to constrain $v\sin i_\star\cos\lambda$ but not $v\sin
i_\star\sin\lambda$. For this reason we chose to parameterize the RM
effect with the quantities $\sqrt{v\sin i_\star} \cos\lambda$ and
$\sqrt{v\sin i_\star} \sin\lambda$, rather than $v\sin i_\star$ and
$\lambda$. The reason for the square roots is to give a constant
Jacobian between the fitting parameters and the ``physical''
parameters $v\sin i_\star$ and $\lambda$. As a result, uniform priors
in our fitting parameters correspond to the desired uniform priors in
$v\sin i_\star$ and $\lambda$. With no square roots, and no other
adjustment to the fitting procedure, the implicit prior would be
linear in $v\sin i_\star$ and would thereby bias the results toward
faster rotation rates.

\begin{table*}[t]
 \begin{center}
  \caption{Parameters of the WASP-1 system\label{tab:wasp1}}
    \smallskip 
       \begin{tabular}{l  r@{\,\,$\pm$\,\,}l    }
          \tableline\tableline
          \hline
	  \noalign{\smallskip}
	  Parameter &  \multicolumn{2}{c}{ Values} \\
	  \noalign{\smallskip}	 
	  \hline
	  \noalign{\smallskip}
          \multicolumn{3}{c}{Parameters mainly derived from photometry} \\
          \noalign{\smallskip}
          \hline
	  \noalign{\smallskip}
          Midtransit time $T_{\rm c}$ [BJD$_{\rm TDB}$$-$2\,400\,000] &  $54461.8630$ & $0.0002$     \\
          Period, $P$ [days]                                                                    &  $2.5199464$   &   $0.0000008$   \\
          $\cos i_{\rm o}$                                                                      &  \multicolumn{2}{c}{$0.000$--$0.034$}   \\   
          Fractional stellar radius,         $R_{\rm \star}/a $                        &  $0.173$  & $^{0.003}_{0.001}$   \\
          Fractional planetary radius,	 $  R_{\rm p}/R_{\star} $              &  $0.1059$ &  $0.0006$  \\
          $u_{1}$+$u_{2}$                                                                        &  $0.20$ &  $0.05$ \\         
          \noalign{\smallskip}
          \hline
          \noalign{\smallskip}
          \multicolumn{3}{c}{Parameters mainly derived from RVs} \\
          \noalign{\smallskip}
          \hline 
          \noalign{\smallskip}
          Velocity offset, HDS [m\,s$^{-1}$]                                           & $0$ & $1.5$ \\
          Velocity offset, HIRES [m\,s$^{-1}$]                                         &   $-17$ & $2$  \\
          Velocity semiamplitude, $K_{\star}$  [m\,s$^{-1}$]                    & $125$ & $5$ \\
         $\sqrt{v \sin i_{\star}} \sin \lambda$ [km\,s$^{-1}$]                 &  $-0.6$ & $0.9$ \\
         $\sqrt{v \sin i_{\star}} \cos \lambda$ [km\,s$^{-1}$]               & $0.31$ & $0.25$  \\
          \noalign{\smallskip}
          \hline
          \noalign{\smallskip}
          \multicolumn{3}{c}{Indirectly derived parameters} \\
	  \noalign{\smallskip}
          \hline
          \noalign{\smallskip}
          Orbital inclination,   $i_{\rm o}$   [$^{\circ}$]             &         \multicolumn{2}{c}{$88$--$92$}    \\
          Full duration, $T_{14}$  [hr]           &     $3.684$ &  $0.017$     \\
          Ingress or egress duration, $T_{12}$  [min]      &  $21.5$ &    $^{0.8}_{0.2}$       \\
          Projected stellar rotation speed, $v \sin i_{\star}$ [km\,s$^{-1}$]     &   $0.7$ & $^{1.4}_{0.5}$            \\
          Projected spin-orbit angle,  $\lambda$    [$^{\circ}$]             &      $-59$ & $^{99}_{26}$ \\
	  \noalign{\smallskip}
	  \tableline
           \noalign{\smallskip}
           \noalign{\smallskip}
     \end{tabular}
   \end{center}
\end{table*}

The other model parameters were a constant RV offset specific to each
spectrograph; the semiamplitude of the star's orbital velocity
($K_\star$), which controls the RV slope that is observed on each
transit night; the orbital period ($P$); a particular time of
midtransit ($T_{\rm c}$); the stellar radius in units of the orbital
distance ($R_\star/a$); the cosine of the orbital inclination ($\cos
i_o$); the planet-to-star radius ratio ($R_p/R_\star$); two quadratic
limb-darkening coefficients $u_1$ and $u_2$ for describing the
$z'$-band photometric data; and a linear limb-darkening coefficient
$u$ to describe the spectroscopic transit (for which most of the
signal is derived from the region 5000--6200~\AA).  According to the
tables of \cite{claret2004}, appropriate choices for the
limb-darkening coefficients are $u_{1}$=0.1666, $u_{2}$=0.3583,
$u=0.6$. We allowed $u_1+u_2$ to be a free parameter and held fixed
$u_1-u_2$ at the tabulated value of $-0.1917$, since the difference is
only weakly constrained by the data (and in turn has little effect on
the other parameters). Likewise we held fixed $u=0.6$.  We assumed the
orbit to be circular, as no sign of any eccentricity was detected by
\cite{cameron2007}, \cite{madhusudhan2009}, \cite{wheatley2010}, or
\cite{pont2011}.\footnote{ In particular, \cite{madhusudhan2009}
    reported an upper limit of $e < 0.088$ with $95.4\%$
    confidence. If the orbit were actually eccentric, in contradiction
    of our modeling assumption, then the main change would be that our
    result for the velocity semiamplitude $K_\star$ would be
    biased. The results for the spin-orbit parameters would not be
    significantly affected.} All of the time stamps of the
  spectroscopic and photometric data were placed on the BJD$_{\rm
    TDB}$ system using the algorithm of \cite{eastman2010}.

The fitting statistic was
{\small
 \begin{eqnarray}
\label{eqn:stat_wasp1}
\chi^{2} & = &
     \sum_{i=1}^{57} \left[\frac{{\rm RV}_i{\rm (o)}-{\rm RV}_i{\rm (c)}}{\sigma_{{\rm RV}, i}}\right]^2 +
   \sum_{j=1}^{1134} \left[\frac{{\rm F}_j{\rm (o)}-{\rm F}_j{\rm (c)}}{\sigma_{{\rm F}, j}}\right]^2 \nonumber\\
  &  &  + \left(\frac{K_{\star} - 115\, {\rm m\,s}^{-1}}{11\, {\rm m\,s}^{-1}}\right)^2 ,
\end{eqnarray}
}
where the first two terms are sums-of-squares over the residuals
between the observed (o) and calculated (c) values of the radial
velocity (RV) and relative flux (F), and the last term represents a
prior constraint on $K_\star$ based on the results of
\cite{cameron2007}. Below we will repeat the analysis including the
constrain on  $v\sin i_\star$ found in Section~\ref{sect:wasp1_obs}.

We solved for the model parameters and their uncertainties using the
Markov Chain Monte Carlo (MCMC) algorithm \citep{tegmark2004}. We used
a chain length of $2\times 10^{6}$ steps and set the size of the steps
in each parameter yielding an acceptance rate of about 30\%. Before
running the chain we increased the uncertainties of the HIRES RVs by
adding a ``stellar jitter'' term of $5$~m\,s$^{-1}$ in quadrature to
the internally-estimated uncertainties. This choice of jitter term
produced a reduced $\chi^{2}$ of unity when that data set was fitted
alone. In making this step we have assumed that the extra RV noise is
well described as Gaussian and uncorrelated. This is consistent with
the appearance of the residuals shown in
Figure~\ref{fig:wasp1_rv}, although we acknowledge there is no
guarantee. Table~\ref{tab:wasp1_rv} reports the original,
internally-estimated uncertainties without any jitter term.

The results for the RM parameters are displayed in
Figure~\ref{fig:wasp1_mcmc}, and the results for all the parameters
are given in Table~\ref{tab:wasp1}.  As anticipated, the weak
detection (or nondetection) of the RM effect led to tighter bounds on
$v\sin i_\star \cos\lambda$ than on $v\sin i_\star \sin\lambda$. This
is why the contours in Figure~\ref{fig:wasp1_mcmc} reach to large
values of $v\sin i_\star$ for small values of $\cos\lambda$ ($\lambda
\approx \pm 90^\circ$).

In an attempt to break the degeneracy between $v\sin i_\star$ and
$\lambda$ we refitted the data with a prior constraint on $v\sin
i_\star$. Based on the results of Section~\ref{sect:wasp1_obs}, we
used a one-sided Gaussian prior, taking the value of unity for $v\sin
i_\star < 2.9$~km~s$^{-1}$ and falling off as a Gaussian function with
$\sigma=0.5$~km\,s$^{-1}$ for higher values. The results from this
more constrained MCMC analysis are shown in
Figure~\ref{fig:wasp1_mcmc}. The modified bounds on $\lambda$ are
$-53\pm^{98}_{29}$\,$^{\circ}$. This analysis disfavors
$\lambda\approx \pm90^\circ$ as this would require larger $v\sin
i_\star$. However it is not possible to tell definitively whether the
positive or negative solution is correct. Within the 95\% confidence
contour, all prograde orbits are allowed.

A different approach is to use a prior constraint on $v$, the actual
rotation speed of the star, based on its spectral type and age.
\cite{schlaufman2010} recently presented a formula for a main-sequence
star's expected rotation period, given its mass and age. He based the
formula on the observed rotation periods of stars in young clusters
along with the \cite{skumanich1972} law $v\propto t^{-1/2}$. He
further showed that this formula gives a good description of the
$v\sin i_\star$ distribution of stars in the SPOCS catalog
\citep{valenti2005}. For WASP-1, he found an expected value $v =
8.6\pm0.5$~km\,s$^{-1}$ where the uncertainty is based only on the
uncertainties in the age and mass of WASP-1, and does not account for
any uncertainty due to intrinsic scatter in the mass-age-period
relation, which seems to be about 3 times larger than the formal
uncertainty [see, e.g., Fig.~3 of \cite{schlaufman2010}]. Taking $v =
8.6\pm 1.5$~km\,s$^{-1}$ together with our result $v\sin i_\star <
2.9$~km\,s$^{-1}$, the implication is $\sin i_\star < 0.34$, i.e., the
star is viewed close to pole-on.

One might wonder if the Skumanich law is really applicable to stars
with close-in planets, which may have undergone significant evolution
due to tidal interactions. For the case of WASP-1, at least, there is
supporting evidence for relatively rapid rotation, based on its
observed color and chromospheric emission. \cite{aigrain2004} explain
how to use a star's observed $B-V$ and $log_{10}R'_{\rm HK}$ indices
to predict its rotation period. Applied to WASP-1, for which
$B-V=0.53$ and $log_{10}R'_{\rm HK} = -5.114$ (Knutson et al.~2010),
we find a rotation period of 12.9~days. Using a stellar radius of
$1.45$~R$_{\odot}$ \citep{charbonneau2007}, the predicted rotation
speed is $v = 5.7$~km\,s$^{-1}$, in good agreement with the value
expected from the statistical analysis by \citet{schlaufman2010}.
 
We therefore have two independent lines of evidence for a high
obliquity, or equivalently, we have strong evidence against the
well-aligned scenario in which $\sin i_\star \approx 1$ and $\lambda
\approx 0^\circ$. (1) The absence of a strong RM effect requires
either that $|\lambda| \approx 90^{\circ}$, or else $v \sin i_{\star}$
is very low ($<$1~km~s$^{-1}$). The latter possibility is incompatible
with a well-aligned star ($\sin i_\star \approx 1$), because the
rotation rate for a star of the given mass and age is expected to be
$8.6\pm 1.5$~km\,s$^{-1}$. The observed color and chromospheric
  activity level also suggest a rotation speed of this order. (2)
Independently of the RM effect, our determination of $v \sin
i_{\star}$ based on the observed width of the spectral lines is much
lower than the value of the expected rotation speed, which implies a
low $\sin i_\star$. In short, it is likely that the stellar and
orbital spins are misaligned along the line of sight, and it is
possible that they are also misaligned within the sky plane.

\subsection{Comparison with previous results}

\cite{simpson2011} reported $\lambda$=$-79^{+4.5}_{-4.3}$\,$^{\circ}$
for WASP-1b, based on observations taken during and after a planetary
transit with the SOPHIE spectrograph on the 1.93m telescope of the
Observatoire de Haute-Provence. Their value for $\lambda$ is
compatible with our result. However their uncertainty is much smaller
than we have found. What causes this difference in obtained confidence
intervals?

Their RV data, reproduced in the bottom panel of
Figure~\ref{fig:wasp1_rv}, appears to have a higher amplitude than was
seen in our data. This could lead to a somewhat higher result for $v
\sin i_{\star}$ but would not by itself affect the very strong
correlation between $v \sin i_{\star}$ and $\lambda$. Rather, the
important differences are in the methods of analysis. There are two
main differences.

Firstly, rather than jointly fitting the photometric and spectroscopic
data as we have done, \cite{simpson2011} fitted their spectroscopic
data using independent Gaussian priors on the photometric parameters
$a/R_{\star}$, $R_{\rm p}/R_{\star}$, and $i_{\rm o}$. The problem is
that those parameters are themselves very strongly correlated and
their posterior distributions are far from Gaussian. In particular
their photometric priors excluded very low impact parameters, while we
find that $b\approx 0$ is allowed. To avoid this problem it is better
to analyze photometric and spectroscopic data together, or to place
priors on the relatively uncorrelated parameters $T_{14}$, $T_{12}$
and $R_{\rm p}/R_{\star}$ \citep{carter2008}.

Secondly, \cite{simpson2011} used a prior on $v\sin i_\star$ based on
the spectroscopic analysis of \cite{stempels2007}, which gave $v\sin
i_{\star} = 5.79\pm0.35$~km\,s$^{-1}$. As explained in Section 3.1 and
shown in Figure~\ref{fig:wasp1_spec}, our spectroscopic analysis
implies a slower projected rotation rate. Their prior on $v\sin
i_\star$ pushed their solution towards higher $v \sin i_{\star}$ and
excluded aligned configurations of the projected axes.

\begin{figure}
  \begin{center}
   \includegraphics[width=8.cm]{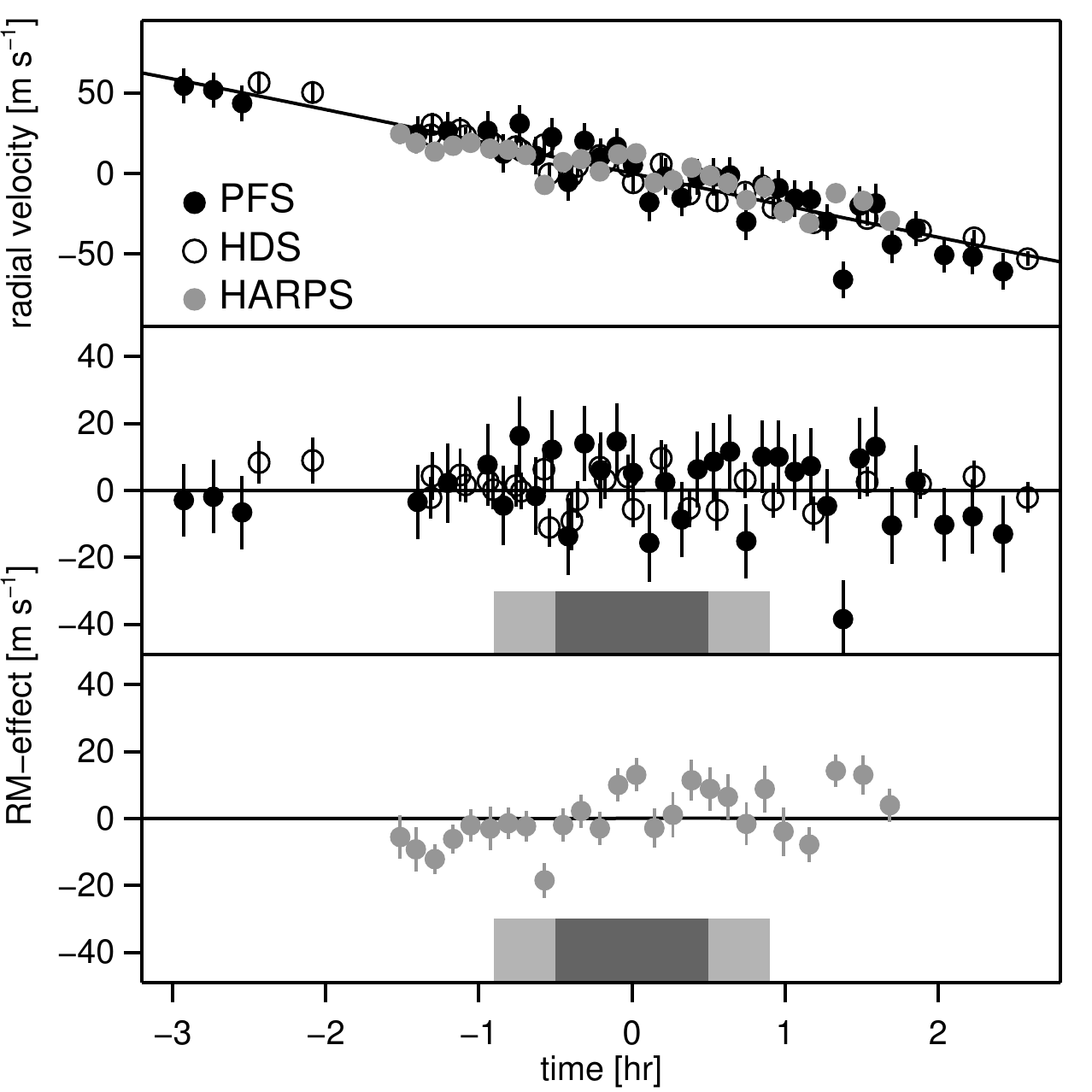}
   \caption {\label{fig:wasp2_rv} {\bf Spectroscopy of WASP-2
       transits. } Similar to Figure~\ref{fig:wasp1_rv}. Black symbols
     are PFS data, and open symbols are HDS data. Gray symbols are
     the HARPS data of \cite{triaud2010}, which are shown for
     comparison but were not used during the fitting process. The
     upper panel shows the data and the best-fitting orbital model. In
     the lower two panels, our best-fitting orbital model has been
     subtracted from the data.}
  \end{center}
\end{figure}

\section{WASP-2}
\label{sect:wasp2}

\begin{figure*}
  \begin{center} \includegraphics[width=15.cm]{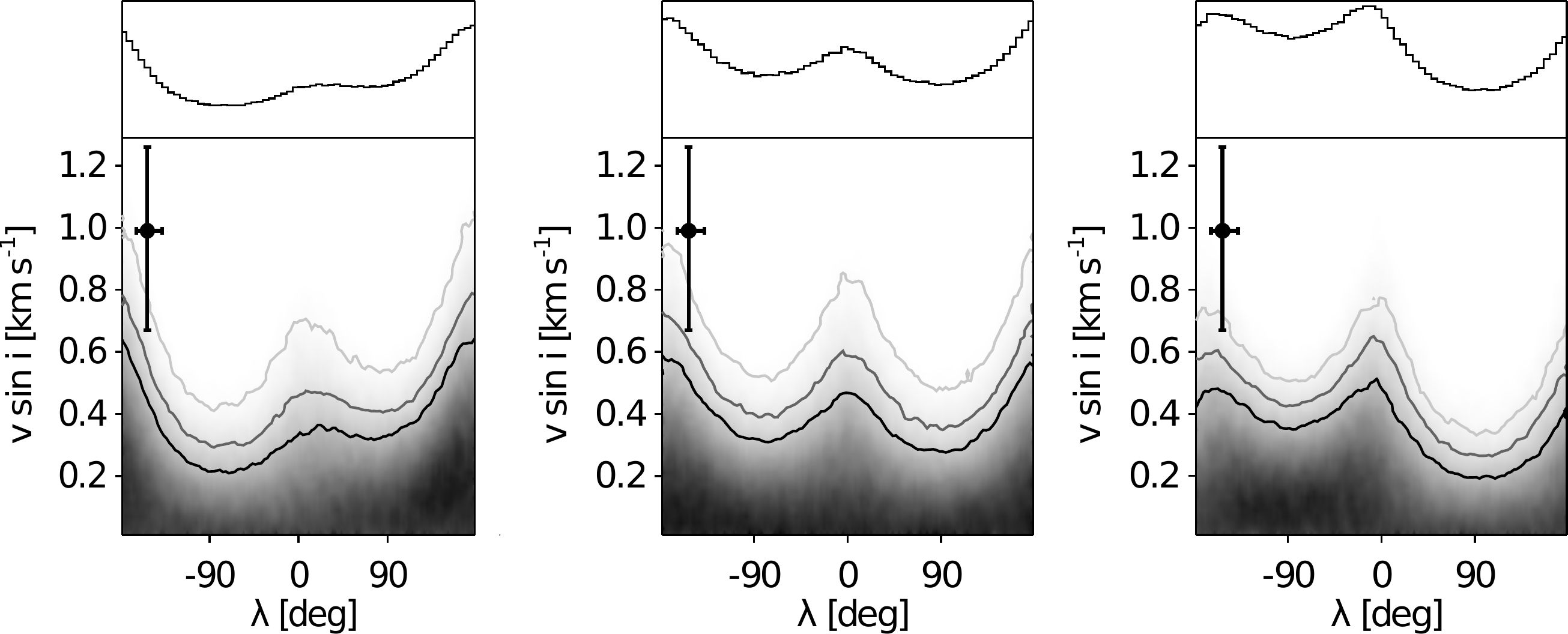}
    \caption {\label{fig:wasp2_mcmc} {\bf Results for $v \sin
        i_{\star}$ and $\lambda$ in the WASP-2 system.} Similar to
      Figure~\ref{fig:wasp1_mcmc}, but for WASP-2. The gray scale
      plots indicate the posterior probability densites, marginalized
      over all other parameters. The contours represent the 2-D
      68.3\%, 95\%, and 99.73\% confidence limits. The one-dimensional
      marginalized distributions for $\lambda$ are on top of the
      contour plot.  The left panels show the results for the MCMC
      analysis with no prior applied to $K_{\star}$, the middle panel
      shows the results with a prior on $K_{\star}$ as shown in
      equation \ref{eqn:stat_wasp2}, and the right panel shows the
      result for the prior with half the confidence interval. The
      black dots with error bars mark the results by
      \cite{triaud2010}. The error bars are those quoted by
        \cite{triaud2010}, representing 68.3\% confidence intervals in
        $\lambda$ and $v \sin i_{\star}$ marginalized over all other
        parameters. They are not strictly appropriate for this
        two-dimensional plot.  We refer the reader to Figure~3 of
        \cite{triaud2010} to view their two-dimensional posterior
        distribution. }
  \end{center}
\end{figure*}

\subsection{Observations and basic stellar parameters}

We conducted spectroscopic observations of WASP-2 transits with the
Magellan (Clay) 6.5\,m telescope and the Subaru 8.2\,m telescope. With
Magellan we used the Planet Finding Spectrograph (PFS;
\citealt{crane2010}) to gather 35 spectra spanning the transit of
2010~August~26/27. With Subaru we used the HDS to obtain 21 spectra
spanning the transit of 2007~September~4/5, and 10 spectra spanning
the transit of 2007~September~19/20. Again we employed the iodine-cell
technique to derive precise radial velocities.  All the RVs are given
in Table~\ref{tab:wasp2_rv}, and plotted in Figure~\ref{fig:wasp2_rv}.
As was the case for WASP-1, we found no clear evidence for the RM
effect.

To check on the basic stellar parameters, we also obtained a
high-quality template spectrum with Keck/HIRES, so that we could use
the same SME-based analysis that was used for WASP-1.  We obtained
$T_{\rm eff} = 5206\pm 50$~K, $\log g=4.51\pm 0.10$, $[M/H] = 0.04\pm
0.05$, and $v\sin i_\star = 1.3\pm 0.5$~km~s$^{-1}$. The assumed
macroturbulent velocity was 3.11~km~s$^{-1}$.  Using the MORPH code
described in Section 3.1, we found that the WASP-2 lines are no
broader than the Solar lines, and estimate $v\sin i_\star \lsim 1.5$
km\,s$^{-1}$.

\subsection{Analysis}

The transit impact parameter for WASP-2b is large, with
\cite{torres2008} having reported
$b=0.724^{+0.017}_{-0.028}$. Therefore, based on the reasoning of
Section 2, the nondetection of the RM effect implies that both $v\sin
i_\star \cos\lambda$ and $v\sin i_\star\sin \lambda$ are small, which
is only possible for low $v\sin i_\star$. Unlike the case for WASP-1b,
the RM effect for WASP-2 cannot be suppressed by having the planet's
trajectory coincide with the sky-projected rotation axis.  We
therefore expect the nondetection to lead to an upper limit on $v\sin
i_\star$ and no information about $\lambda$.

\begin{table*}[t]
 \begin{center}
  \caption{Parameters of the WASP-2 system\label{tab:wasp2}}
    \smallskip 
       \begin{tabular}{l  r@{\,\,$\pm$\,\,}l    }
          \tableline\tableline
          \hline
	  \noalign{\smallskip}
	  Parameter &  \multicolumn{2}{c}{ Values} \\
	  \noalign{\smallskip}	 
	  \hline
	  \noalign{\smallskip}
          \multicolumn{3}{c}{Parameters mainly controlled by prior knowledge} \\
          \noalign{\smallskip}
          \hline
	  \noalign{\smallskip}
          Midtransit time $T_{\rm c}$ [BJD$_{\rm TDB}$$-$2\,400\,000] & $53991.51530$ & $0.00017$     \\
          Period, $P$ [days]                                                                    &   $2.15222144$   &  $0.00000040$   \\
          $\cos i_{\rm o}$                                                                             &   $0.091$ & $0.007$  \\   
          Fractional stellar radius,         $R_{\rm \star}/a $                        &  $0.125$  & $0.005$   \\
          Fractional planetary radius,	 $  R_{\rm p}/R_{\star} $              &  $0.1309$ &  $0.0015$  \\  
          \noalign{\smallskip}
          \hline
          \noalign{\smallskip}
          \multicolumn{3}{c}{Parameters mainly derived from RVs} \\
          \noalign{\smallskip}
          \hline 
          \noalign{\smallskip}
          Velocity offset, PFS [m\,s$^{-1}$]                                           & $-2$ & $2$ \\
          Velocity offset, HIRES [m\,s$^{-1}$]                                         &   $-23.6$ & $2$  \\
          Velocity semiamplitude, $K_{\star}$  [m\,s$^{-1}$]                    & $164$ & $4$ \\
         $\sqrt{v \sin i_{\star}} \sin \lambda$ [km\,s$^{-1}$]           &  $-0.02$  & $0.28$ \\
         $\sqrt{v \sin i_{\star}} \cos \lambda$ [km\,s$^{-1}$]               & $-0.038$ & $0.36$  \\
          \noalign{\smallskip}
          \hline
          \noalign{\smallskip}
          \multicolumn{3}{c}{Indirectly derived parameters} \\
	  \noalign{\smallskip}
          \hline
          \noalign{\smallskip}
          Orbital inclination,   $i_{\rm o}$   [$^{\circ}$]                                    &  $84.8$ & $0.5$  \\
          Full duration, $T_{14}$  [hr]           &  $1.799$ &  $0.037$     \\
          Ingress or egress duration, $T_{12}$  [min]      &  $24.2$ &    $2.4$       \\
          Projected stellar rotation speed, $v \sin i_{\star}$ [km\,s$^{-1}$]     &   \multicolumn{2}{c}{$<$0.5 (2$\sigma$)}              \\
          Projected spin-orbit angle,  $\lambda$    [$^{\circ}$]          &     \multicolumn{2}{c}{ all values allowed}   \\
	  \noalign{\smallskip}
	  \tableline
           \noalign{\smallskip}
           \noalign{\smallskip}
     \end{tabular}
   \end{center}
\end{table*}

For the quantitative analysis our procedure was similar to that used
for WASP-1. The RVs were modeled as sum of contributions from a
circular orbit, the RM effect, and a constant offset specific to each
spectrograph. We used a prior on $K_\star$ from \cite{triaud2010}, but
with a doubled uncertainty (see below), and also tested the
sensitivity of the results to this prior as described below. Since the
photometric parameters are already precisely determined and we do not
have any new photometric data, we implemented priors on the full
transit duration ($T_{14}$), the ingress or egress duration
($T_{12}$), the radius ratio ($R_p/R_\star$) from
\cite{charbonneau2007}, and the transit ephemeris based on the
analysis of \cite{southworth2010}.  The fitting statistic was
{\small
\begin{eqnarray}
\label{eqn:stat_wasp2}
\chi^{2} & = & \sum_{i=1}^{66} 
\left[\frac{{\rm RV}_i{\rm (o)}-{\rm RV}_i{\rm (c)}}{\sigma_{{\rm RV},i}}\right]^2\nonumber\\   
&  &  + \left(\frac{T_{\rm c, BJD} - 2453991.51530}{0.00017}\right)^2 +\left(\frac{P- 2\fd15222144}{0\fd00000039}\right)^2  \nonumber\\         
&  &  + \left(\frac{T_{14} - 1.799\,{\rm hr}}{0.0035\,{\rm hr}}\right)^2 +
\left(\frac{T_{12} - 24.6\,{\rm min}}{2.4\,{\rm min}}\right)^2 \nonumber\\ 
&  &  + \left(\frac{{R_{\rm p}/R}_{\star} - 0.1309}{ 0.0015}\right)^2  + \left(\frac{K_{\star} - 153.6\, {\rm m\,s}^{-1}}{6\, {\rm m\,s}^{-1}}\right)^2 ,
\end{eqnarray}
}
where the symbols have the same meaning as in section
\ref{sect:wasp1}. For the PFS data, a ``stellar jitter'' term of
10~m\,s$^{-1}$ was added in quadrature to the internally-estimated
uncertainties to give a reduced $\chi^2$ of unity. This probably
reflects the limitations of the current algorithm that is used to
estimate uncertainties, which is geared toward much brighter stars.

Our results are presented in Table~\ref{tab:wasp2} and are illustrated
by the contours in the middle panel of
Figure~\ref{fig:wasp2_mcmc}. [The single solid point in
Figure~\ref{fig:wasp2_mcmc} represents the result of
\cite{triaud2010}, which will be discussed below.] As expected, $v\sin
i_\star$ is constrained to low values but $\lambda$ can assume any
value from $-180^\circ$ to $+180^\circ$.

The three different panels of Figure~\ref{fig:wasp2_mcmc} show the
results of different choices for the prior on $K_\star$. We wondered
about the sensitivity of the results to this prior because the star is
a late-type star and might be expected to have starspots, which can
cause the observed RV slope surrounding the transit phase to be
steeper than one would expect from the spectroscopic orbital
parameters. Starspots always move across the stellar disk from the
approaching limb to the receding limb, and thereby produce an RM-like
effect with a negative slope, which is added to the actual orbital
velocity gradient. This effect can be seen in a number of RM datasets
presented in the literature, most notably for the highly spotted star
CoRoT-2 \citep{bouchy2008}. Depending on the distribution of
measurements before, during and after transit this might introduce
different biases in the results for $\lambda$ and $v\sin i_\star$.

In Figure~\ref{fig:wasp2_mcmc}, the left panel shows the results with
no prior on $K_\star$, the middle panel shows the result for a prior
on $K_\star$ as in Eqn.~(\ref{eqn:stat_wasp2}), and the right panel
employed the same prior but with a width of 3~m~s$^{-1}$ instead of
6~m~s$^{-1}$. Evidently the results are not very sensitive to the
prior on $K_\star$: in all cases $v\sin i_\star$ must be low and
$\lambda$ may have any value. For concreteness our final results given
in Table~\ref{tab:wasp2} are based on a prior with a width of
6~m~s$^{-1}$ (i.e., the analysis depicted in the middle panel).

One interesting feature of Figure~\ref{fig:wasp2_mcmc} is that the
posterior probability density for $\lambda$ has peaks near $0^\circ$
and 180$^\circ$. For these choices of $\lambda$, larger values of
$v\sin i_\star$ are compatible with the nondetection. This is a
general result when fitting RM data with a low signal-to-noise ratio
of a high-$b$ system, and can be understood as follows. For $\lambda$
near $0^\circ$ and 180$^\circ$, the RM signal is antisymmetric about
the midtransit time. In such cases $v\sin i_\star$ and $K_\star$ are
strongly correlated parameters, since small changes in either
parameter produce changes to the RM signal that are antisymmetric
about the midtransit time. This leads to larger confidence intervals
for $v\sin i_\star$. In contrast, for $\lambda = \pm 90^\circ$ the RM
signal is symmetric about the midtransit time; it is a pure redshift
or blueshift. Here, the parameters $K_\star$ and $v\sin i_\star$ are
uncorrelated and the allowed region for $v\sin i_\star$ shrinks. To
put it another way: by fitting for the systemic velocity and
$K_\star$, we have effectively applied a high-pass filter to the RV
data, and thereby reduced the amplitude of any RM signal with $\lambda
= \pm 90^\circ$ in comparison to the higher-frequency signal that is
produced with $\lambda$ near $0^\circ$ and 180$^\circ$. This causes
the allowed range of $v\sin i_\star$ to be higher for $\lambda$ near
$0^\circ$ and 180$^\circ$. This explanation was confirmed with further
numerical experiments described in Section \ref{sect:wasp2_compare}.

As with WASP-1, one may try to gain more information on the spin orbit
alignment by using prior constraints on $v\sin i_\star$ or $v$, but in
this case not much refinement is possible.  The analysis of the WASP-2
template spectrum gives an upper limit $v\sin i_\star \lsim
1.5$~km~s$^{-1}$ which is not constraining in this context. Also,
there have been no reports of photometric variations due to star
spots, and hence no stellar rotation period has been
determined. Likewise, \cite{schlaufman2010} found that the expected
rotation speed for this system, based on its mass and age, is
$1.61$~km\,s$^{-1}$ with an uncertainty range of 1.72~km\,s$^{-1}$
(presumably an asymmetric error interval). Because of the large
uncertainty it is not possible to draw any conclusion about $\sin
i_\star$, and for this reason \cite{schlaufman2010} did not identify
WASP-2 as a probable case of a misaligned star.

As an additional check on the expected stellar rotation speed, we used
the approach of \cite{aigrain2004} to estimate the rotation period of
WASP-2, as we did for WASP-1. In this case, $B-V=0.84$ and
$log_{10}R'_{\rm HK} = -5.054$ (Knutson et al.\ 2010), from which we
derive a stellar rotation period of $46$~days. Together with an
stellar radius of $0.81$~R$_{\odot}$ \citep{charbonneau2007}, this
gives a rotation speed of $v = 0.9$~km\,s$^{-1}$ which is in line with
the low speed predicted by \cite{schlaufman2010}.

\subsection{Comparison with previous results}
\label{sect:wasp2_compare}

A transit of WASP-2 was observed by \cite{triaud2010} with the HARPS
spectrograph. Their data are shown in the bottom panel of
Figure~\ref{fig:wasp2_rv}. Based on the HARPS data they found
$\lambda=-153^{+15}_{-11}$~degrees (a retrograde orbit) and $v \sin
i_{\star}$ = $0.99^{+0.27}_{-0.32}$~km\,s$^{-1}$. Our data are not
compatible with those parameters. When we fixed $\lambda$ and $v \sin
i_{\star}$ at the values found by these researchers, and refitted our
data, the minimum $\chi^{2}$ rose from $60.9$ to $72.6$, giving
$\Delta \chi^{2} = 11.7$. What can have caused the difference between
our results and theirs?

\cite{daemgen2009} found that WASP-2 has a neighboring star (a
companion or chance alignment) at an angular separation of
$0.7$~arcsec, close enough to have been possibly included within the
spectrograph slit or fiber in some cases. It is hard to predict the
exact effect that the additional starlight would have on the
spectroscopic analysis, but as the neighbor is 4~mag fainter than
WASP-2, and as its spectral type and systemic velocity are likely
quite different from that of WASP-2, we consider it unlikely that
variable contamination by this star is responsible for the differing
results. We are therefore led to look elsewhere for an explanation.

\begin{figure}
  \begin{center} 
    \includegraphics[width=8cm]{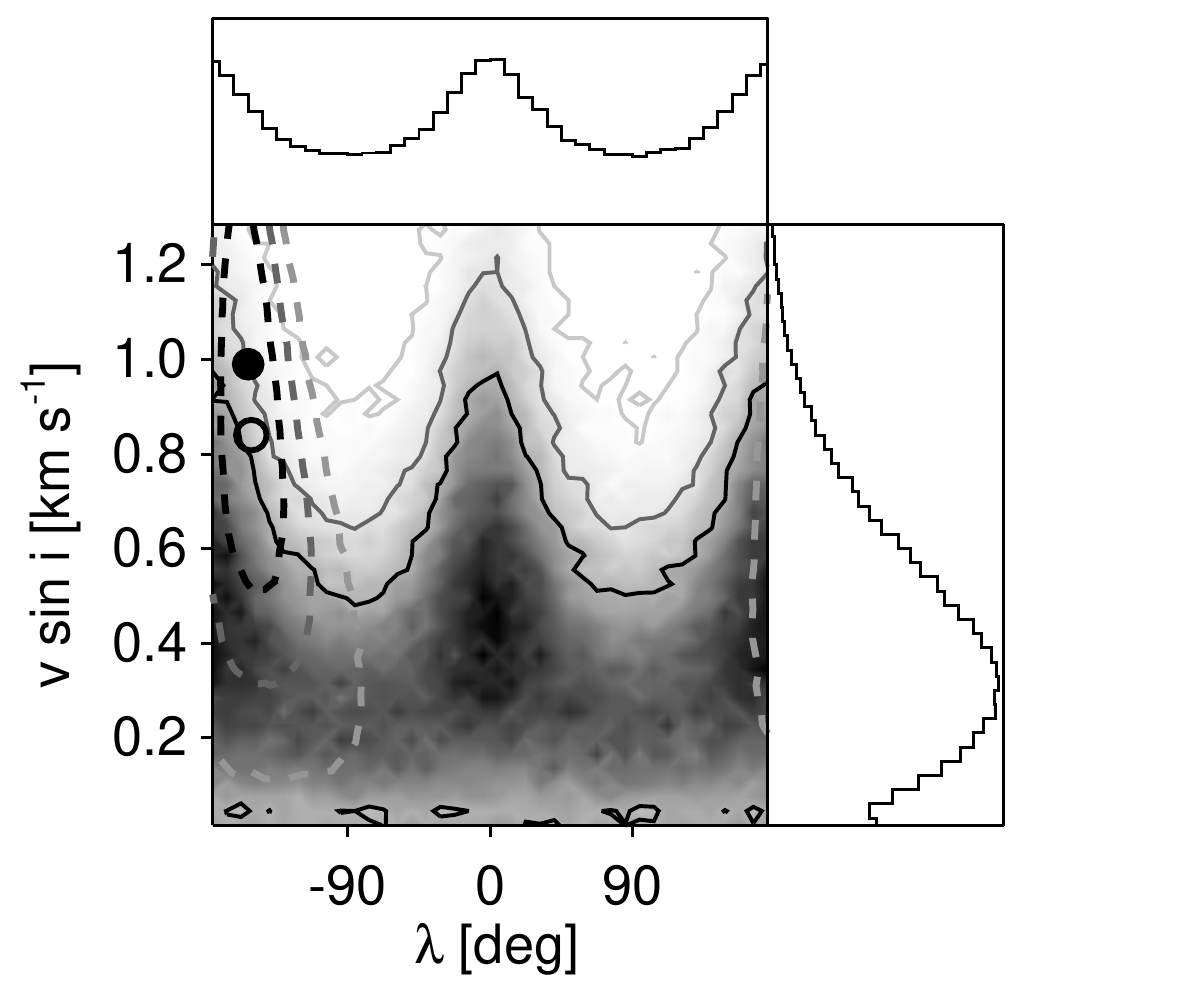}
    \caption {\label{fig:mook_data} {\bf Results for simulated data
        sets with no RM effect.} Similar to Figure 9, but this time
      based on the analysis of $2\times10^5$ simulated data sets with no RM effect
      but with the same time sampling  and roughly the same RV precision
      as the HARPS data.  The gray shades show the density of the best-fitting values of
      $v\sin i_\star$ and $\lambda$. The contours enclose 68.3\%,
      95\%, and 99.73\% of the best-fitting values. The solid circle shows the
      \cite{triaud2010} result. The open circle and the dashed
      contours show our results of fitting the actual HARPS dataset
      with our MCMC routine.}
  \end{center}
\end{figure}

One relevant difference in the analysis procedures is that
\cite{triaud2010} used uniform priors in $v \sin i_{\star} \sin
\lambda $ and $v \sin i_{\star} \cos \lambda$, thereby adopting a
prior that is linear in $v\sin i_\star$. This is in contrast to our
prior which was uniform in $v\sin i_\star$. Their prior pushes $v \sin
i_{\star}$ to higher values and therefore pushes $\lambda$ near
$0^{\circ}$ or $180^{\circ}$ (see Figure~\ref{fig:wasp2_mcmc}). When
we refitted their data using our procedure, we found a lower $v\sin
i_\star$ and an enlarged confidence interval, as expected. The open
circle and the thick dashed lines in Figure~\ref{fig:mook_data}
represent our fit to the HARPS data. However this difference in priors
cannot explain the entire discrepancy: even our reanalysis of the
HARPS data gives $\lambda=-151^{+20}_{-13}$~degrees and $v \sin
i_{\star}$ = $0.84\pm0.35$~km\,s$^{-1}$.

This apparently statistically significant result is surprising since
the RM effect is not apparent by visual inspection of the data
(Figure~\ref{fig:wasp2_rv}). The data during the transit does not
appear too different from the data outside of the transit. If the RM
effect had been measured but not modeled, then one would expect the
residuals between the data and the best-fitting orbital model would
have a higher scatter inside the transit than outside the transit. For
our data this is not the case. For the HARPS data set the rms residual
of the out-of-transit data is 6.9\,m\,s$^{-1}$, as compared to
7.2\,m\,s$^{-1}$ during transit. This represents only a marginal
increase in scatter.

This led us to conduct some numerical experiments on fitting random
noise with similar characteristics to the HARPS data. We used the
timestamps of the HARPS transit-night data, and simulated RV data
based on only the best-fitting orbital model for WASP-2. We added
Gaussian ``measurement'' uncertainties with a standard deviation of
7.0\,m\,s$^{-1}$. Then we fitted this mock dataset together with the
photometric priors using a Levenberg-Marquardt least-squares
minimization routine. This was repeated $2\times10^5$ times with different
realizations of the measurement errors.\footnote{We did not use the
  MCMC algorithm as it would take to long to make chains for $10^5$
  data sets, and because we are only interested in the best fitting
  values of $v\sin i_\star$ and $\lambda$ for each mock data set and
  not the individual confidence intervals.} The density distribution
of the $2\times10^5$ best-fitting solutions is shown in
Figure~\ref{fig:mook_data}. As discussed in Section 4.2, we found that
even though the mock data had no RM effect at all, there is a clear
tendency to ``find'' solutions near $\lambda = 0^{\circ}$ or
$180^{\circ}$. This should raise a concern about the claimed detection
of the RM effect with $\lambda\approx 0^\circ$ or $\approx$180$^\circ$
with a low signal-to-noise ratio. The result of our fitting code
applied to the actual HARPS data (open circle and dashed contours in
Figure~\ref{fig:mook_data}) gives values for $v\sin i_\star$ and
$\lambda$ that are within the area containing 95\% of the mock-data
solutions. In this sense the ``false alarm'' probability (the odds of
finding such an apparently significant retrograde orbit when fitting
only random noise) is at least 5\%. It is probably higher, when one
considers that the true noise may not be uncorrelated and Gaussian. We
therefore conclude that the current data do not provide secure
information on the orientation of the stellar spin relative to the
orbital spin.

\section{Discussion and summary}
\label{sect:discussion}

We have presented two nondetections of the RM effect for the
transiting planets WASP-1b and WASP-2b. In both cases we gathered
high-resolution, high--signal-to-noise ratio spectra on nights
spanning transits, using multiple large telescopes. For WASP-1 there
is a weak indication of a prograde RM effect, and for WASP-2 we did
not detect the RM effect. Due to the differences in the transit
geometry, and in the stellar type, we arrived at different conclusions
about the relative orientation of the stellar spin and orbit in each
case.

Because the transit of WASP-1b has a very low impact parameter, the
only way to produce a low-amplitude RM effect is to have nearly
perpendicular sky projections of the spin and orbital axes (implying a
large misalignment in the sky plane), or to have a very low $v\sin
i_\star$. The latter option also implies a likely misalignment,
because the resulting upper limit on $v\sin i_\star$ is lower than the
expected $v$ for a star of the given age and mass.  A similar
comparison can be made between the expected $v$ and the lower $v\sin
i_\star$ that is estimated from the breadth of spectral absorption
lines. Thus the data give strong evidence for misalignment, although
it is not certain whether the misalignment is mainly along the line of
sight, or in the sky plane, or both.

For WASP-2b, no information on $\lambda$ was gained from our
nondetection, mainly because this star is expected to be a slow
rotator. The upper limit on $v\sin i_\star$ from the RM nondetection
is within the expected range of $v$ for a star of the given mass and
age. An analysis of previous HARPS data favored a retrograde orbit for
the planet, but we have argued that this may have been a statistical
false alarm. Numerical experiments confirm that fitting random noise
with an RM model can produce false detections with nearly the same
amplitude as the claimed detection. For a firmer conclusion one would
need to gather more spectroscopic data during transits. These same
numerical experiments should lead to a re-evaluation of other cases in
which the RM effect was detected with low statistical significance,
such as TrES-2b \citep{Winn2008}.

We now put these results into the context of the pattern noted by
\citet{winn2010} and \citet{schlaufman2010}, that hot stars tend to
have high obliquity. The proposed boundary line between ``hot'' and
``cool'' star was around $T_{\rm eff} = 6250$~K.

For WASP-2, \cite{cameron2007} measured an effective temperature of
$5200\pm 200$~K, and from our HIRES spectrum we found $5206\pm 50$~K.
Thus there is consensus that WASP-2 is a cool star. The finding of a
retrograde orbit by \cite{triaud2010} was a strong exception to the
proposed pattern. Our data and our analysis led us to conclude that
the spin-orbit angle for this system is undetermined, and therefore
that WASP-2 is not an exception.

For WASP-1, \cite{cameron2007} measured an effective temperature of
$6200\pm 200$~K. Further observations and spectroscopic analysis were
presented by \cite{stempels2007}, who found $T_{\rm eff} = 6110\pm
45$~K. Our analysis of a HIRES spectrum gave $T_{\rm eff} = 6213\pm
51$~K, or 100~K hotter than the determination by
\cite{stempels2007}. Probably the reason for the difference is that
\cite{stempels2007} used the H$\alpha$ line profile as the main
constraint on $T_{\rm eff}$, while our analysis used the standard SME
wavelength intervals which exclude H$\alpha$ \cite[][Table
3]{valenti2005}. It is beyond the scope of this article to evaluate
the relative merits of these different methods for establishing an
accurate effective temperature scale. Instead we note that the
SME-based scale that we have used is similar or identical to the scale
that has been used for the other transit-hosting stars, and therefore
the scale on which the proposed boundary of 6250~K is relevant. In
this light it seems that WASP-1, with $T_{\rm eff}$~(SME)~$=6213\pm
51$~K is very near the boundary. Therefore the finding of a high
obliquity neither corroborates nor weakens the proposed pattern,
although WASP-1 may serve as a useful point in establishing the
sharpness of the transition from mainly-misaligned to mainly-aligned.

\acknowledgments We thank G. Marcy and M. Holman for help gathering
some of the data presented here. We are grateful to the anonymous
referee for a prompt and insightful report, and to Amaury Triaud for
comments on the manuscript. S.A.\ acknowledges support by a
Rubicon fellowship from the Netherlands Organization for Scientific
Research (NWO). J.N.W.\ acknowledges support from a NASA Origins grant
(NNX09AD36G). This research has made use of the Simbad database
located at {\tt http://simbad.u-strasbg.fr/}.

{\it Facilities:} 
\facility{Keck}.
\facility{Subaru}.
\facility{Magellan}.

\appendix

\begin{table}
  \caption{Relative Radial Velocity measurements of WASP-1}
  \label{tab:wasp1_rv}
  \begin{center}
    \smallskip
    \begin{tabular}{l c c c}
      \hline
      \hline
      \noalign{\smallskip}
      Time [BJD$_{\rm TDB}$] & RV~[m~s$^{-1}$] & Unc.~[m~s$^{-1}$] & Spectrograph \\
      \noalign{\smallskip}
      \hline
      $  2454345.83725$  &  $     33.86$  &  $   2.82$ & HIRES \\
      $  2454345.84471$  &  $     26.32$  &  $   2.57$ & HIRES \\
      $  2454345.85916$  &  $     29.23$  &  $   2.40$ & HIRES \\
      $  2454345.86449$  &  $     19.53$  &  $   2.94$ & HIRES \\
      $  2454345.86988$  &  $     16.26$  &  $   2.76$ & HIRES \\
      $  2454345.87525$  &  $     28.59$  &  $   2.71$ & HIRES \\
      $  2454345.88059$  &  $     26.28$  &  $   2.75$ & HIRES \\
      $  2454345.88603$  &  $     18.84$  &  $   2.76$ & HIRES \\
      $  2454345.89146$  &  $     27.23$  &  $   2.63$ & HIRES \\
      $  2454345.89684$  &  $     18.61$  &  $   3.05$ & HIRES \\
      $  2454345.90219$  &  $     23.13$  &  $   2.87$ & HIRES \\
      $  2454345.90758$  &  $     13.82$  &  $   2.92$ & HIRES \\
      $  2454345.91296$  &  $     18.70$  &  $   2.84$ & HIRES \\
      $  2454345.91830$  &  $     19.45$  &  $   2.65$ & HIRES \\
      $  2454345.92368$  &  $     -0.07$  &  $   2.78$ & HIRES \\
      $  2454345.92907$  &  $      7.22$  &  $   2.77$ & HIRES \\
      $  2454345.93453$  &  $     -8.89$  &  $   2.73$ & HIRES \\
      $  2454345.93990$  &  $      2.32$  &  $   2.74$ & HIRES \\
      $  2454345.94841$  &  $      1.12$  &  $   2.81$ & HIRES \\
      $  2454345.95376$  &  $     -1.71$  &  $   2.91$ & HIRES \\
      $  2454345.95912$  &  $    -11.69$  &  $   2.60$ & HIRES \\
      $  2454345.96926$  &  $     -5.30$  &  $   2.73$ & HIRES \\
      $  2454345.97461$  &  $    -11.05$  &  $   2.85$ & HIRES \\
      $  2454345.97998$  &  $    -12.33$  &  $   3.32$ & HIRES \\
      $  2454345.98537$  &  $    -10.78$  &  $   3.19$ & HIRES \\
      $  2454345.99861$  &  $    -15.13$  &  $   2.93$ & HIRES \\
      $  2454346.00397$  &  $    -18.29$  &  $   2.98$ & HIRES \\
      $  2454346.00936$  &  $    -12.28$  &  $   3.03$ & HIRES \\
      $  2454346.01468$  &  $    -19.91$  &  $   2.99$ & HIRES \\
      $  2454346.02006$  &  $    -16.84$  &  $   2.99$ & HIRES \\
      $  2454346.03551$  &  $    -22.71$  &  $   3.01$ & HIRES \\
      $  2454346.06736$  &  $    -33.51$  &  $   2.97$ & HIRES \\
      $  2454346.13628$  &  $    -62.64$  &  $   2.96$ & HIRES \\
      $  2454346.14171$  &  $    -64.20$  &  $   2.78$ & HIRES \\
      $  2454318.09458$  &  $     63.61$  &  $   9.84$ & HDS \\
      $  2454318.12329$  &  $     48.93$  &  $   9.86$ & HDS \\
      $  2454318.13785$  &  $     36.74$  &  $  10.64$ & HDS \\
      $  2454350.88327$  &  $     51.29$  &  $   5.81$ & HDS \\
      $  2454350.89783$  &  $     45.61$  &  $   5.81$ & HDS \\
      $  2454350.90899$  &  $     41.88$  &  $   6.36$ & HDS \\
      $  2454350.91661$  &  $     31.89$  &  $   6.08$ & HDS \\
      $  2454350.92423$  &  $     44.93$  &  $   6.19$ & HDS \\
      $  2454350.93185$  &  $     32.72$  &  $   6.31$ & HDS \\
      $  2454350.93945$  &  $     40.32$  &  $   6.57$ & HDS \\
      $  2454350.94707$  &  $     35.60$  &  $   7.06$ & HDS \\
      $  2454350.95469$  &  $     30.06$  &  $   5.77$ & HDS \\
      $  2454350.96231$  &  $     24.87$  &  $   6.22$ & HDS \\
      $  2454350.96992$  &  $     32.04$  &  $   6.13$ & HDS \\
      $  2454350.97753$  &  $     21.08$  &  $   5.92$ & HDS \\
      $  2454351.03561$  &  $      6.90$  &  $   9.11$ & HDS \\
      $  2454351.05624$  &  $     -5.33$  &  $   7.87$ & HDS \\
      $  2454351.06385$  &  $     -5.00$  &  $   7.65$ & HDS \\
      $  2454351.07146$  &  $    -17.63$  &  $   8.04$ & HDS \\
      $  2454351.08247$  &  $    -15.36$  &  $   6.90$ & HDS \\
      $  2454351.09703$  &  $    -21.98$  &  $   6.42$ & HDS \\
      $  2454351.11159$  &  $    -20.25$  &  $   8.19$ & HDS \\
      $  2454351.12615$  &  $    -29.56$  &  $   7.36$ & HDS \\
      \noalign{\smallskip}
      \hline
    \end{tabular}  
  \end{center}
\end{table}

\begin{table}
  \caption{Relative Radial Velocity measurements of WASP-2}
  \label{tab:wasp2_rv}
  \begin{center}
    \smallskip
    \begin{tabular}{l c c c}
      \hline
      \hline
      \noalign{\smallskip}
      Time [BJD$_{\rm TDB}$] & RV~[m~s$^{-1}$] & Unc.~[m~s$^{-1}$] & Spectrograph \\
      \noalign{\smallskip}
      \hline
      $  2454348.72936$  &  $     47.57$  &  $   6.15$ & HDS \\
      $  2454348.73875$  &  $     46.93$  &  $   5.29$ & HDS \\
      $  2454348.74635$  &  $     41.88$  &  $   5.75$ & HDS \\
      $  2454348.75397$  &  $     37.82$  &  $   5.51$ & HDS \\
      $  2454348.76158$  &  $     23.31$  &  $   5.67$ & HDS \\
      $  2454348.76920$  &  $     28.01$  &  $   5.53$ & HDS \\
      $  2454348.77680$  &  $     30.28$  &  $   5.66$ & HDS \\
      $  2454348.78442$  &  $     17.81$  &  $   5.26$ & HDS \\
      $  2454348.79204$  &  $     29.38$  &  $   5.52$ & HDS \\
      $  2454348.79965$  &  $     10.77$  &  $   5.73$ & HDS \\
      $  2454348.80728$  &  $      6.57$  &  $   6.10$ & HDS \\
      $  2454348.81489$  &  $     11.95$  &  $   5.42$ & HDS \\
      $  2454348.82250$  &  $      2.32$  &  $   5.23$ & HDS \\
      $  2454348.83352$  &  $     -6.93$  &  $   5.20$ & HDS \\
      $  2454348.84809$  &  $     -4.37$  &  $   4.45$ & HDS \\
      $  2454348.86264$  &  $    -11.82$  &  $   4.52$ & HDS \\
      $  2454348.87720$  &  $    -16.39$  &  $   4.75$ & HDS \\
      $  2454348.89175$  &  $    -29.35$  &  $   4.59$ & HDS \\
      $  2454348.90632$  &  $    -33.89$  &  $   4.52$ & HDS \\
      $  2454348.92089$  &  $    -44.93$  &  $   4.18$ & HDS \\
      $  2454348.95000$  &  $    -54.24$  &  $   4.86$ & HDS \\
      $  2454363.74812$  &  $     79.94$  &  $   6.36$ & HDS \\
      $  2454363.76268$  &  $     73.84$  &  $   6.94$ & HDS \\
      $  2454363.79521$  &  $     53.93$  &  $   7.07$ & HDS \\
      $  2454363.80283$  &  $     50.60$  &  $   7.88$ & HDS \\
      $  2454363.81044$  &  $     45.11$  &  $   7.11$ & HDS \\
      $  2454363.81804$  &  $     40.19$  &  $   6.31$ & HDS \\
      $  2454363.82566$  &  $     41.40$  &  $   7.46$ & HDS \\
      $  2454363.83326$  &  $     22.31$  &  $   8.60$ & HDS \\
      $  2454363.84088$  &  $     34.76$  &  $   7.03$ & HDS \\
      $  2454363.84850$  &  $     28.17$  &  $   6.82$ & HDS \\
      $  2455435.53391$  &  $     56.65$  &  $   4.38$ & PFS \\
      $  2455435.54192$  &  $     54.05$  &  $   4.56$ & PFS \\
      $  2455435.54978$  &  $     45.81$  &  $   4.58$ & PFS \\
      $  2455435.59753$  &  $     26.70$  &  $   4.62$ & PFS \\
      $  2455435.60576$  &  $     28.47$  &  $   6.46$ & PFS \\
      $  2455435.61657$  &  $     28.78$  &  $   7.18$ & PFS \\
      $  2455435.62090$  &  $     14.48$  &  $   6.44$ & PFS \\
      $  2455435.62533$  &  $     33.21$  &  $   6.00$ & PFS \\
      $  2455435.62971$  &  $     13.14$  &  $   6.00$ & PFS \\
      $  2455435.63414$  &  $     24.86$  &  $   6.09$ & PFS \\
      $  2455435.63854$  &  $     -3.11$  &  $   5.62$ & PFS \\
      $  2455435.64295$  &  $     22.52$  &  $   5.21$ & PFS \\
      $  2455435.64731$  &  $     12.40$  &  $   5.33$ & PFS \\
      $  2455435.65172$  &  $     18.86$  &  $   5.71$ & PFS \\
      $  2455435.65618$  &  $      7.36$  &  $   5.93$ & PFS \\
      $  2455435.66062$  &  $    -15.67$  &  $   5.78$ & PFS \\
      $  2455435.66496$  &  $      0.39$  &  $   5.01$ & PFS \\
      $  2455435.66941$  &  $    -12.93$  &  $   5.32$ & PFS \\
      $  2455435.67376$  &  $      0.00$  &  $   5.24$ & PFS \\
      $  2455435.67817$  &  $      0.19$  &  $   5.95$ & PFS \\
      $  2455435.68261$  &  $      1.08$  &  $   4.92$ & PFS \\
      $  2455435.68702$  &  $    -27.79$  &  $   4.90$ & PFS \\
      $  2455435.69140$  &  $     -4.66$  &  $   4.44$ & PFS \\
      $  2455435.69580$  &  $     -6.84$  &  $   4.64$ & PFS \\
      $  2455435.70025$  &  $    -13.39$  &  $   5.39$ & PFS \\
      $  2455435.70460$  &  $    -13.75$  &  $   5.15$ & PFS \\
      $  2455435.70901$  &  $    -27.76$  &  $   4.88$ & PFS \\
      $  2455435.71338$  &  $    -63.64$  &  $   5.72$ & PFS \\
      $  2455435.71787$  &  $    -17.76$  &  $   6.67$ & PFS \\
      $  2455435.72222$  &  $    -16.32$  &  $   6.81$ & PFS \\
      $  2455435.72666$  &  $    -41.92$  &  $   5.87$ & PFS \\
      $  2455435.73313$  &  $    -31.84$  &  $   4.42$ & PFS \\
      $  2455435.74091$  &  $    -48.37$  &  $   4.57$ & PFS \\
      $  2455435.74864$  &  $    -49.44$  &  $   4.97$ & PFS \\
      $  2455435.75692$  &  $    -58.54$  &  $   5.51$ & PFS \\
       \noalign{\smallskip}
	\hline
      \end{tabular}  
    \end{center}
  \end{table}

\end{document}